\documentclass[aps,prc]{revtex4}

\usepackage{amsmath,amssymb,amsopn}
\usepackage{graphicx}
\usepackage{subfigure}
\newcommand{\be}{\begin{equation}}
\newcommand{\ee}{\end{equation}}
\newcommand{\ba}{\begin{eqnarray}}
\newcommand{\ea}{\end{eqnarray}}
\newcommand{\ban}{\begin{eqnarray*}}
\newcommand{\ean}{\end{eqnarray*}}
\newcommand \nn {\nonumber}

\begin{document}

\title{Energy Loss in Unstable Quark-Gluon Plasma}

\author{Margaret E. Carrington\footnote{e-address: carrington@brandonu.ca}}

\affiliation{Department of Physics, Brandon University, Brandon, Manitoba, Canada}
\affiliation{Winnipeg Institute for Theoretical Physics, Winnipeg, Manitoba, Canada}

\author{Katarzyna Deja\footnote{e-address: kdeja@fuw.edu.pl}}
\affiliation{National Centre for Nuclear Research, Warsaw, Poland}

\author{Stanis\l aw Mr\' owczy\' nski\footnote{e-address: mrow@fuw.edu.pl}}

\affiliation{Institute of Physics, Jan Kochanowski University, Kielce, Poland}
\affiliation{National Centre for Nuclear Research, Warsaw, Poland}

\date{October 8, 2015}

\begin{abstract}

The momentum distribution of quark-gluon plasma at the early stage of a relativistic heavy-ion collision is anisotropic and consequently the system, which is assumed to be weakly coupled, is unstable due to chromomagnetic plasma modes. We consider a high-energy parton which flies across such an unstable plasma, and the energy transfer between the parton and the medium is studied as an initial value problem. In the case of equilibrium plasmas, the well-known formula of collisional energy loss is reproduced. The unstable plasma case is much more complex, and the parton can lose or gain energy depending on the initial conditions. The extremely prolate and extremely oblate systems are considered as examples of unstable plasmas, and two classes of initial conditions are discussed. When the initial chromodynamic field is uncorrelated with the color state of the parton, it typically looses energy, and the magnitude of the energy loss is comparable to that in an equilibrium plasma of the same density. When the initial chromodynamic field is induced by the parton, it can be either accelerated or decelerated depending on the relative phase factor. With a correlated initial condition, the energy transfer grows exponentially in time and its magnitude can much exceed the absolute value of energy loss in an equilibrium plasma. The energy transfer is also strongly directionally dependent. Consequences of our findings for the phenomenology of jet quenching in relativistic heavy-ion collisions are briefly discussed.

\end{abstract}

\maketitle

\section{Introduction}

The observation of jet quenching in central collisions of relativistic heavy ions is considered to be a signal that quark-gluon plasma  (QGP) is produced, because only a medium with deconfined color charges could stop a multi-GeV parton within a few femtometers, see {\it e.g.} the reviews \cite{Peigne:2008wu,Majumder:2010qh}.  The energy loss of a high energy parton is a key element of a quantitative understanding of the jet quenching  phenomenon and has been intensively studied for two decades, but the problem is far from being completely solved  \cite{Peigne:2008wu,Majumder:2010qh}. 

QGP is produced in the early stage of a relativistic heavy-ion collision and spends most of its lifetime in a state of local equilibrium.
The energy loss of a high-energy parton is therefore conventionally computed in a locally equilibrated plasma which evolves hydrodynamically. However, a QGP reaches a state of local equilibrium only after a short but finite time interval \cite{Heinz:2004pj,Bozek:2010aj}, and the momentum distribution of pre-equilibrium plasma is anisotropic. Even if the anistropic phase is very short lived, it might have a significant effect on the energy loss of a test parton because the weakly coupled anisotropic QGP is unstable due to spontaneously growing chromomagnetic modes (for a review see \cite{Mrowczynski:2005ki}). The anisotropic plasma is populated with large chromodynamic fields which will strongly influence the test parton. Therefore, the energy loss of the parton in the brief anisotropic phase might constitute a significant fraction of the total energy loss responsible for the experimentally observed jet quenching. An analysis of this system is very complicated however, because the unstable QGP evolves quickly due to the presence of unstable modes, and the energy loss calculation has to be treated as an initial value problem. Our results show that the energy loss is strongly time dependent, and this dependence is much stronger than the switching-on effect studied in \cite{Peigne:2005rk,Adil:2006ei}. In contrast to the weakly coupled plasma studied here, energy loss in the strong coupling regime computed in the framework of AdS/CFT duality is rather similar in equilibrium and in far-from-equilibrium plasma at the same energy density \cite{Chesler:2013cqa}. 

In this paper we study a highly energetic parton which gains or loses energy through interactions with the chromodynamic fields present in the QGP. The effect of elastic interactions is called {\it collisional energy loss} and that of gluon emission {\it radiative energy loss}. In the case of light partons - light quarks and gluons -  radiative energy loss is expected to give the dominant contribution to the total energy loss \cite{Peigne:2008wu,Majumder:2010qh}. For heavy quarks the radiative energy loss is presumably less important due to the so-called dead cone effect \cite{Peigne:2008wu,Majumder:2010qh}. Collisional energy loss in anisotropic QGP was studied in Ref. \cite{Romatschke:2004au} but the unstable plasma was treated as a static medium and the interaction of the test parton with exponentially growing chromodynamic fields was ignored. The energy loss computed in \cite{Romatschke:2004au} thus misses the key feature of unstable plasmas. 

We will treat the test parton as a classical particle with ${\rm SU}(N_c)$ color charge, the dynamics of which is described by the Wong equations \cite{Wong:1970fu} combined with the linearized Yang-Mills equations. We assume that the typical momenta of the collective modes are much less than the typical momenta of the plasma constituents. This approach is equivalent to using QCD within the hard-loop (HL) approximation \cite{Litim:2001db}. In the equilibrium limit the time dependence of the energy loss disappears and we reproduce the soft part of the collisional energy loss \cite{Bjorken:1982tu,Thoma:1990fm,Braaten:1991we,Mrowczynski:1991da}, where the momentum transfer is of the order of the Debye mass.  The energy loss due to soft interactions diverges logarithmically with the upper limit of the momentum transfer, which we call $k_{\rm max}$. In our anisotropic calculations, we also find an approximately logarithmic dependence on $k_{\rm max}$. This ultraviolet sensitivity is expected since the approach is classical, and signals the necessity to combine the classical contribution to the energy loss at small wave vectors with the quantum contribution at higher ones. A quantum approach to parton energy loss in unstable plasma needs to be developed. 

Our crucial finding is that depending on the initial conditions the test parton can either lose or gain energy when traversing the unstable QGP. In an equilibrium plasma the parton loses energy and the energy change per unit length $dE/dx$ is negative. If the parton gains energy from the plasma fields,  $dE/dx$ is positive. Although the energy transfer can be either negative or positive, depending on the situation, we frequently use the term `energy loss' generically to describe both cases.  Our results show that when the intial conditions are chosen in a certain way, the magnitude of the energy loss increases exponentially, which indicates that the unstable modes play an important role. At late enough times, the energy loss can be much bigger than in equilibrium plasmas. It is also strongly directionally dependent.

The acceleration of a test particle in a plasma system might seem rather exotic, but the phenomenon is well known in the physics of electromagnetic plasmas. It is caused by the electric field associated with plasma waves in the system. Charged particles are carried forward on the electrostatic wave with a motion like surfing with speed equal to the phase velocity of the wave, and can therefore be boosted to very high energies. This picture motivates the idea to use a plasma excited by a laser or particle beam as a particle accelerator. A mechanism was proposed in 1979 by Tajima and Dawson \cite{Tajima:1979bn}, and was experimentally verified soon afterwards \cite{Joshi:1984}. Since plasmas can sustain accelerating fields orders of magnitude larger than those in the radio-frequency modules of standard accelerators, small plasma devices can be extremely efficient. In the experiment in Ref. \cite{Leemans:2006dx}, electrons were accelerated to an energy as high as 1 GeV  over a distance of 3.3 cm, demonstrating immense promise for affordable and compact plasma accelerators for various applications.
 
This paper is organized as follows. In Sec.~\ref{sec-form-general} we derive the general energy loss formula of a relativistic classical parton in an unstable QCD plasma which depends on the initial conditions. The method is similar to the approach developed earlier by one of us to study the spectra of chromodynamic fluctuations \cite{Mrowczynski:2008ae} and the momentum broadening of a fast parton  \cite{Majumder:2009cf}. The equilibrium limit is discussed in Sec.~\ref{sec-stable} where we show that the dependence on the initial conditions drops out, and our expression reduces to the familiar equilibrium result. In Sec.~\ref{sec-ini-cond} we introduce the two classes of initial conditions that we will apply to unstable plasmas, and in Sec.~\ref{sec-vacuum} we study the effect of self-interaction which needs to be subtracted from the energy loss formula. We develop our formalism in Sec.~\ref{sec-aniso-plamsas}, and in Sec.~\ref{sec-ex-pro-obla} we apply it to extremely prolate and oblate systems.  Our results are summarized and the outlook is discussed in Sec.~\ref{sec-conclusions}. 
In Appendix~\ref{sec-reality-EL} we show that the energy loss is real, and the temporal axial gauge is compared with the Feynman-Lorentz gauge in Appendix \ref{sec-gauge}.

A preliminary account of our findings was published in the series of conference proceedings \cite{Carrington:2011uj,Carrington:2012hv,Carrington:2013tz,Carrington:2014yra}. In these reports, we used a specific choice of initial conditions and we were not aware that our results crucially depend on this choice, to the extent that the test parton can not only loose energy in an unstable plasma, but could also gain the energy, depending on how the initial conditions are chosen. Therefore, the results presented in \cite{Carrington:2011uj,Carrington:2012hv,Carrington:2013tz,Carrington:2014yra} do not reveal the full complexity of the problem.

Throughout the paper we use natural units where $\hbar = c =1$. The indices $i,j,k = 1, 2, 3$ and $\mu, \nu = 0, 1, 2, 3$ label, respectively, the Cartesian spatial coordinates and those of Minkowski space.

\section{General formula}
\label{sec-form-general}

Our formalism is based on the HL QCD effective action. It can be shown that the Wong equations \cite{Wong:1970fu} and the linearized Yang-Mills (Maxwell) equations can be obtained directly from this action \cite{Litim:2001db}. The Wong equations describe the motion of a classical parton moving through the fields of a plasma. The motion of the parton changes the field configurations, which is self-consistently taken into account through the linearized Yang-Mills equations which relate the chromodynamic fields to the parton charge and current. We emphasize that even though the Yang-Mills equations are linearized by the HL approximation, HL QCD is {\it not} equivalent to HL QED (up to an overall factor), because the gluons contribute to the color charge density and current in these equations. 

The Wong equations, which describe the motion of a parton in a chromodynamic field, are usually written in the Lorentz covariant form \cite{Wong:1970fu}
\ba
\label{EOM-1a}
\frac{d x^\mu(\tau)}{d \tau} &=& u^\mu(\tau ) ,
\\
\label{EOM-1b}
\frac{d p^\mu(\tau)}{d \tau} &=& g Q^a(\tau ) \, F_a^{\mu \nu}\big(x(\tau )\big) \, u_\nu(\tau ) ,
\\
\label{EOM-1c}
\frac{d Q_a(\tau)}{d \tau} &=& - g f^{abc} u_\mu (\tau ) \, A^\mu _b \big(x(\tau )\big) \, Q_c(\tau) ,
\ea
where $\tau$, $x^\mu(\tau )$, $u^\mu(\tau)$ and  $p^\mu(\tau)$ are, respectively, the parton's  proper time, trajectory, four-velocity and  four-momentum; $F_a^{\mu \nu}$ and $A_a^\mu$ denote, respectively, the chromodynamic field strength tensor and four-potential in the adjoint representation of the ${\rm SU}(N_c)$ gauge group with the color index $a = 1, \; 2, \dots N_c^2 -1$; $g$ is the coupling constant, which is assumed to be small, and finally $gQ^a$ is the classical color charge of the parton.

The Wong equations (\ref{EOM-1a}), (\ref{EOM-1b}), and (\ref{EOM-1c}) are supplemented by the linearized Yang-Mills equations describing the self-consistent generation of  the chromodynamic field. We write the linearized Yang-Mills equations in a non-covariant three-vector notation where they have the familiar form of Maxwell equations in a medium. In Heaviside-Lorentz electromagnetic units, which are usually used in quantum field theory, we have 
\ba
\label{Maxwell-eqs-x-lin-1}
\nabla \cdot {\bf D}_a(t, {\bf r}) &=& \rho_a (t, {\bf r})
\;,\;\;\;\;\;\;\;\;\;\;\;\;\;
\nabla \cdot {\bf B}_a(t, {\bf r}) = 0 , 
\\[2mm] 
\label{Maxwell-eqs-x-lin-2}
\nabla \times {\bf E}_a(t, {\bf r}) &=& 
- \frac{\partial {\bf B}_a(t, {\bf r})}{\partial t} ,
\;\;\;\;\;
\nabla \times {\bf B}_a(t, {\bf r}) =
{\bf j}_a(t, {\bf r}) +\frac{\partial {\bf D}_a(t, {\bf r})}{\partial t},
\ea
where ${\bf E}_a$, ${\bf D}_a$, ${\bf B}_a$ are the chromoelectric field, chromoelectric induction and chromomagnetic field; $\rho_a$ and ${\bf j}_a$ are the density and current of the test parton, respectively. To close the system of Maxwell equations  (\ref{Maxwell-eqs-x-lin-1}) and (\ref{Maxwell-eqs-x-lin-2}), the  chromoelectric induction is expressed through the chromoelectric field
\be
\label{D-vs-E-x}
D^i_a (t,{\bf r}) = \int dt' \,d^3r' \varepsilon^{ij} (t-t',{\bf r} - {\bf r}') E^j_a (t',{\bf r}') ,
\ee
where $\varepsilon^{ij}(t,{\bf r})$ is the chromodielectric permeability. We note that when the medium is on average color neutral, the chromodielectric permeability is proportional to $\delta^{ab}$, which can be factored out to produce an expression that carries no color indices as in Eq.~(\ref{D-vs-E-x}). We also note that a quantity which depends on color indicies through $\delta^{ab}$, or carries no color indices, is gauge independent. 

To solve the Wong equations (\ref{EOM-1a}), (\ref{EOM-1b}), and (\ref{EOM-1c}) we adopt two simplifying assumptions. The first is that we  choose the gauge condition
\ba
\label{gaugeCondition}
u_\mu (\tau ) \, A^\mu _a \big(x(\tau )\big) = 0 ,
\ea
which requires that the potential vanishes along the parton's trajectory. Using this gauge, the third Wong equation (\ref{EOM-1c}) simply states that the parton's charge is a constant of motion, or that $Q_a$ is independent of $\tau$. The second important simplification comes from the fact that we consider a highly energetic parton and assume that its velocity  ${\bf v}$ is constant and ${\bf v}^2=1$. In an equilibrium plasma the characteristic momentum transfer $|\Delta {\bf p}|$ is of order $gT$ and the parton's momentum $|{\bf p}| \gg T$, where $T$ is the temperature. The hard loop approach requires $gT \ll T$, and therefore $|\Delta {\bf v}| \sim |\Delta {\bf p}| / |{\bf p}|\ll 1$. When we consider anisotropic systems, we assume the same hierarchy of scales which gives $|\Delta {\bf v}|\ll 1$. The physical picture is that due to interaction with the chromodynamic field the parton's energy and momentum evolve in time without changing the magnitude of its velocity.

Replacing the proper time $\tau$ by the time $t=\gamma\tau$ and writing $x^i(t) = r^i(t)$ and $u^i(t) = \gamma \,v^i$, the first Wong equation (\ref{EOM-1a}) gives ${\bf r}(t) = {\bf v} t$. Using this result, we obtain from the second Wong equation (\ref{EOM-1b}) with $\mu = 0$ 
\be
\label{e-loss-1}
\frac{dE(t)}{dt} = g Q^a \,{\bf E}_a \big( t,{\bf r}(t) \big) \cdot {\bf v} .
\ee
Since the current generated by the moving parton equals
\be
\label{current-x}
{\bf j}_a(t,{\bf r}) = g Q^a {\bf v} \delta^{(3)}({\bf r} - {\bf v}t) ,
\ee
we rewrite Eq.~(\ref{e-loss-1}) as
\be
\label{e-loss-2}
\frac{dE(t)}{dt} = \int d^3r \, {\bf E}_a(t,{\bf r})\cdot {\bf j}_a(t,{\bf r})  .
\ee
To obtain the energy loss we must solve equations (\ref{Maxwell-eqs-x-lin-1}) and (\ref{Maxwell-eqs-x-lin-2}) for the electric field and substitute into equation (\ref{e-loss-2}). 

The electric field that appears in equations (\ref{e-loss-1}) and (\ref{e-loss-2}) is the total electric field, which is the sum of the external field generated directly by the moving test parton  and the induced electric field produced by the charge distributions and currents that are induced by the parton in the plasma medium. The external electric field gives the parton's self-interaction and does not contribute to the energy loss. The energy loss comes physically from the motion of the parton into the opposing induced electric field. We derive below an expression for the total electric field from Maxwell's equations. At the end of the procedure, we must either show that the self-interaction does not contribute to the energy loss, or we must subtract it. 

It seems clear from Eq.~(\ref{e-loss-2}) that if the parton moves into an electric field of opposite orientation to its current, the change in the energy will be negative and we have energy loss.  We will show, however, that is not always the case. If the calculation is done as an initial value problem, then the sign of the energy transfer crucially depends on the choice of initial conditions. 

To solve Maxwell's equations we use the usual method which is to Fourier transform the differential equations to change them into algebraic equations which can be easily solved. However, we do not use a standard (two-sided) Fourier transform. Our problem is to track the evolution of a parton starting from some arbitrary initial time (which we take to be $t=0$) and calculate its behavior at future times.  The non-equilibrated plasma is not time-translation invariant, and the energy loss formula should depend on the initial conditions, which means that we need to formulate the calculation as an initial value problem. In order to do this, we use a one-sided Fourier transformation defined as
\ba
\label{1side-forward}
f(\omega,{\bf k}) &=& \int_0^\infty dt \int d^3r
e^{i(\omega t - {\bf k}\cdot {\bf r})}
f(t,{\bf r}) ,
\\[2mm]
\label{1side-inverse}
f(t,{\bf r}) &=& \int_{-\infty +i\sigma}^{\infty +i\sigma}
\frac{d\omega}{2\pi} \int \frac{d^3k}{(2\pi)^3}
e^{-i(\omega t - {\bf k}\cdot {\bf r})} f(\omega,{\bf k}) .
\ea
The inverse transformation (\ref{1side-inverse}) involves the real parameter $\sigma > 0$ which is chosen so that the integral over $\omega$ is taken along a straight line in the complex $\omega$-plane, parallel to the real axis and above all singularities of $f(\omega,{\bf k})$. 

The one-sided Fourier transform of the current (\ref{current-x}) is obtained from Eq.~(\ref{1side-forward}) where the time integral is defined through the limit
\be
\lim_{\epsilon \rightarrow 0^+}
\int_0^\infty dt \,e^{i(\omega - {\bf k}\cdot {\bf v} +i 0^+)t}
= \frac{i}{\omega - {\bf k}\cdot {\bf v}+i 0^+} ,
\ee
which gives
\be
\label{current-k}
{\bf j}_a(\omega,{\bf k}) = 
\frac{i g Q^a {\bf v}}{\omega - {\bf k}\cdot {\bf v}+i 0^+} .
\ee
This procedure is mathematically equivalent to multiplying the current in equation (\ref{current-x}) by a factor $e^{- 0^+ t}$, which can be interpreted physically as imposing the boundary condition that the current goes to zero as the time approaches infinity. 

The one-sided Fourier transform of the relation (\ref{D-vs-E-x}) provides 
\be
\label{D-vs-E}
D^i_a(\omega, {\bf k}) = \varepsilon^{ij}(\omega, {\bf k})
E^j_a(\omega, {\bf k}) .
\ee
Applying the one-sided Fourier transform to the Maxwell equations (\ref{Maxwell-eqs-x-lin-1}) and (\ref{Maxwell-eqs-x-lin-2}) and using the relation (\ref{D-vs-E}) gives
\ba
\label{Maxwell-eqs-k-1}
&& i k^i \varepsilon^{ij}(\omega, {\bf k}) E^j_a(\omega, {\bf k}) = \rho_a (\omega,{\bf k}) ,  
\\[2mm]
\label{Maxwell-eqs-k-2}
&& i k^i B^i_a (\omega,{\bf k}) = 0 ,
\\[2mm]
\label{Maxwell-eqs-k-3}
&& i \epsilon^{ijk} k^j E_a^k(\omega,{\bf k}) = i\omega B_a^i (\omega,{\bf k}) + B_{0a}^i({\bf k}) ,
\\[2mm]
\label{Maxwell-eqs-k-4}
&& i \epsilon^{ijk} k^j B_a^k (\omega,{\bf k})
= j_a^i(\omega,{\bf k})
-i\omega \varepsilon^{ij} (\omega,{\bf k})
E_a^j(\omega,{\bf k}) - D_{0a}^i({\bf k}) ,
\ea
where we have written $B^i_{0a}({\bf k}) \equiv B^i_a(t=0,{\bf k})$ and similarly for $D^i_{0a}({\bf k})$. These initial values come from the time integrals of the time derivatives of fields after performing an integration by parts.  The algebraic equations (\ref{Maxwell-eqs-k-1}) - (\ref{Maxwell-eqs-k-4}) are solved for the field $E^i_a(\omega, {\bf k})$ 
\be
\label{E-field-k}
E^i_a(\omega, {\bf k}) = -i
(\Sigma^{-1})^{ij}(\omega,{\bf k})
\Big[ \omega j_a^j(\omega,{\bf k})
+ \epsilon^{jkl} k^k B_{0a}^l ({\bf k})
- \omega D_{0a}^j ({\bf k}) \Big] ,
\ee
where we have defined the matrix
\be
\label{matrix-sigma}
\Sigma^{ij}(\omega,{\bf k}) \equiv
- {\bf k}^2 \delta^{ij} + k^ik^j
+ \omega^2 \varepsilon^{ij}(\omega,{\bf k}) .
\ee

In a quantum field theory formulation, one uses the retarded gluon polarization tensor $\Pi^{i j}(\omega,{\bf k})$ instead of the dielectric tensor $\varepsilon^{ij}(\omega,{\bf k})$ and the two quantities are related to each other as
\be
\label{diel-tensor}
\varepsilon^{ij} (\omega,{\bf k}) = \delta^{ij} - \frac{1}{\omega^2} \, \Pi^{i j}(\omega,{\bf k}) .
\ee
The full polarization tensor carries Lorentz indices $(\mu, \nu = 0, 1, 2, 3)$, which label coordinates in Minkowski space, and not Cartesian indices $(i,j = 1, 2, 3)$. The components of the polarization tensor not determined by Eq.~(\ref{diel-tensor}) can be reconstructed from the transversality condition $k_\mu \Pi^{\mu \nu}(k) = 0$ with $k^\mu =(\omega,{\bf k})$ which is required by gauge invariance. Using the relation (\ref{diel-tensor}), the matrix (\ref{matrix-sigma}) can be written
\ba
\label{prop}
\Sigma^{ij}(\omega,{\bf k})  = \delta^{ij}(\omega^2-{\bf k}^2)+k^i k^j -\Pi^{ij}(\omega,{\bf k}) 
= (\Delta^{-1}(\omega,{\bf k}))^{ij} .
\ea
This result is recognized as the inverse retarded gluon propagator in the temporal axial gauge ($A^0=0$), and we will write the propagator as $\Delta^{ij}(\omega,{\bf k})$. Although the matrix $\Sigma$ has been derived with no reference to a gauge potential, the form of the gluon propagator depends on the chosen gauge. We return to this alleged conflict in Appendix \ref{sec-gauge}, where we also show that the temporal axial gauge is particularly convenient for our energy loss calculation because it naturally provides gauge independent results. To reach this goal in the Feynman-Lorentz gauge, which, in particular, was used in  \cite{Adil:2006ei}, current conservation must be explicitly enforced.

The effects of the medium are contained in the dielectric tensor, or the polarization tensor. Performing a linear response analysis of kinetic equations in the collisionless limit, or equivalently working in the diagrammatic hard loop approximation, the dielectric tensor of a locally colorless anisotropic plasma equals \cite{Mrowczynski:2000ed,Mrowczynski:2004kv}
\be
\label{eij}
\varepsilon^{ij}(\omega,{\bf k}) = \delta^{ij} +
\frac{g^2}{2\omega} \int {d^3p \over (2\pi)^3} \,
\frac{v^i_p}{\omega - {\bf v}_p\cdot {\bf k}+i0^+}
\Big(\big(1-\frac{{\bf k}\cdot {\bf v}_p}{\omega}\big) \delta^{jk}
+ \frac{v^j_p k^k}{\omega} \Big) \nabla_p^k f({\bf p}) ,
\ee
where ${\bf p}$ and ${\bf v}_p \equiv {\bf p}/|{\bf p}|$ are the momentum and velocity of a {\em massless} parton, and $f({\bf p})$ is the distribution function for hard partons in the plasma. For the ${\rm SU}(N_c)$ gauge group $f({\bf p})= n({\bf p})+ \bar n({\bf p}) +2N_c n_g({\bf p})$, where $n({\bf p})$, $\bar n({\bf p})$, $n_g({\bf p})$ are the distribution functions of quarks, antiquarks and gluons of a single color component. We remind the reader that the chromodielectric tensor does not carry any color indices, as the state corresponding to the momentum distribution $f({\bf p})$ is assumed to be colorless. The $i0^+$ prescription makes the Fourier transformed dielectric tensor $\varepsilon^{ij}(t,{\bf r})$ vanish for $t<0$. In kinetic theory, the infinitesimal quantity  $i0^+$ can be treated as a remnant of inter-particle collisions. Integrating by parts, the chromodielectric tensor (\ref{eij}) can be rewritten in the form
\be
\label{eij-1}
\varepsilon^{ij}(\omega,{\bf k}) = \delta^{ij} -
\frac{g^2}{2\omega^2} \int {d^3p \over (2\pi)^3} \,
\frac{f({\bf p})}{|{\bf p}|}
\bigg[\delta^{ij} +
\frac{k^i v^j_p + v^i_p k^j}
{\omega - {\bf v}_p\cdot {\bf k}+i0^+}
+ \frac{({\bf k}^2 - \omega^2) v^i_p v^j_p}
{(\omega - {\bf v}_p\cdot {\bf k}+i0^+)^2} \bigg] ,
\ee
which is often more convenient to use than the expression (\ref{eij}).  

The energy loss in equation (\ref{e-loss-2}) can now be written in terms of the Fourier transformed field and current. Performing the inverse transformation (\ref{1side-inverse}), we have
\be
\label{e-loss-3}
\frac{dE(t)}{dt} =
\int_{-\infty +i\sigma}^{\infty +i\sigma}
\frac{d\omega}{2\pi}
\int_{-\infty +i\sigma'}^{\infty +i\sigma'}
\frac{d\omega'}{2\pi}
\int {d^3k \over (2\pi)^3}
e^{-i(\omega + \omega')t} \;
{\bf E}_a(\omega,{\bf k}) \,\cdot\,{\bf j}_a(\omega',-{\bf k}) ,
\ee
and substituting  the formulas (\ref{current-k}) and (\ref{E-field-k}) into Eq.~(\ref{e-loss-3}), one obtains
\ba
\label{e-loss-5}
\frac{dE(t)}{dt} &=& -i
\int_{-\infty +i\sigma}^{\infty +i\sigma}
{d\omega \over 2\pi}
\int_{-\infty +i\sigma'}^{\infty +i\sigma'}
{d\omega' \over 2\pi}
\int {d^3k \over (2\pi)^3}
e^{-i(\omega + \omega')t}
\\ [2mm] \nn
&\times&
\frac{i g Q^a v^i}{\omega' + {\bf k}\cdot {\bf v}}
(\Sigma^{-1})^{ij}(\omega,{\bf k})
\Big[ 
\frac{i \omega g Q^a v^j}{\omega - {\bf k}\cdot {\bf v}}
+ \epsilon^{jkl} k^k B_{0a}^l ({\bf k})
- \omega D_{0a}^j ({\bf k}) \Big] .
\ea
The integral over $\omega'$ can be done easily since the integrand has only one pole at $\omega' = - \bar{\omega} \equiv - {\bf k} \cdot {\bf v}$. The result of integration over $\omega'$ is
\ba
\label{e-loss-main}
\frac{dE(t)}{dt} = g Q^a v^i 
\int {d^3k \over (2\pi)^3}
\int_{-\infty +i\sigma}^{\infty +i\sigma} {d\omega \over 2\pi i}
e^{-i(\omega -\bar{\omega})t}
(\Sigma^{-1})^{ij}(\omega,{\bf k})
\Big[ 
\frac{i \omega g Q^a v^j}{\omega - \bar{\omega}}
+ \epsilon^{jkl} k^k B_{0a}^l ({\bf k})
- \omega D_{0a}^j ({\bf k}) \Big], 
\ea
which is the main result of this section. Equation (\ref{e-loss-main}) gives the change of energy of the parton as a function of time, and the expression depends on the initial conditions. The integral over $\omega$ is controlled by the poles of the matrix $\Sigma^{-1}(\omega,{\bf k})$ (or equivalently the gluon propagator) which determine the gluon collective modes in the system. These modes are found as solutions of the dispersion equation 
\be
\label{dis-eq-general}
{\rm det}[\Sigma(\omega,{\bf k})] =0.
\ee
Physically this means that the test parton does not interact with plasma constituents but rather with the plasma collective modes. 

In Sec.~\ref{sec-ini-cond} we discuss how to choose the initial conditions which enter the energy loss formula (\ref{e-loss-main}). In the next section we show that in the equilibrium limit Eq.~(\ref{e-loss-main}) reduces to the familiar result which is independent of the initial conditions.  

\section{Equilibrium limit}
\label{sec-stable}

When the plasma is in equilibrium all collective modes are damped and all poles of the propagator $\Delta^{ij}(\omega,{\bf k}) \equiv (\Sigma^{-1})^{ij}(\omega,{\bf k})$ are located in the lower half-plane of complex $\omega$. The corresponding contributions to the energy loss (\ref{e-loss-main}) exponentially decay in time, and the only stationary contribution is given by the pole $\omega = \bar{\omega} = {\bf k}\cdot {\bf v}$ which comes from the current of the test parton. This means that the terms in Eq.~(\ref{e-loss-main}) which include the initial values of the fields can be neglected. It is mathematically equivalent to use a two-sided Fourier transform from the beginning of the calculation, which means that the initial conditions do not appear in the Maxwell equations (\ref{Maxwell-eqs-k-1}), and the Fourier transform of the current (\ref{current-x}) is just proportional to $\delta(\omega - {\bf k}\cdot {\bf v})$.
The result is that, once again, the only contribution to the integral over $\omega$ comes from $\omega = \bar{\omega} \equiv {\bf k}\cdot {\bf v}$. In both approaches the result is that the energy loss of a high-energy parton traversing an equilibrium plasma is given by the time independent expression
\ba
\label{e-loss-stable}
\frac{dE}{dt} = - ig^2 Q^a Q^a 
v^i v^j
\int {d^3k \over (2\pi)^3} \; \bar{\omega} \;
 (\Sigma^{-1})^{ij}(\bar{\omega},{\bf k}) . 
\ea

Since the parton's color charge is not an observable quantity because of its gauge dependence, the energy loss (\ref{e-loss-stable}) has to be averaged over the parton's color state. This is achieved by means of the relations  
\be
\label{ave-color-charge-1}
\int dQ \,Q_a = 0, 
\ee
and 
\be
\label{ave-color-charge-2}
\int dQ \,Q_a Q_a = C_2 ,
\ee
which are derived in \cite{Litim:2001db}; $C_2=1/2$ for a quark in the fundamental representation of the ${\rm SU}(N_c)$ gauge group and $C_2=N_c$ for a gluon in the adjoint representation. Using the relation (\ref{ave-color-charge-2}), the color averaged energy loss is 
\ba
\label{e-loss-stable-ave}
\frac{d \overline{E}}{dt} &=& -i g^2 C_R v^i v^j
\int {d^3k \over (2\pi)^3} \; \bar{\omega} \;
 (\Sigma^{-1})^{ij}(\bar{\omega},{\bf k}) ,
\ea
where the color factor $C_R$ is given as
\begin{displaymath}
C_R \equiv \left\{
\begin{array}{ccl}
\frac{C_2(N_c^2-1)}{N_c}= \frac{N_c^2-1}{2N_c}
&{\rm for} & {\rm quark} ,  
\\[3mm]
C_2 = N_c &{\rm for}  & {\rm gluon}.
\end{array}
\right.
\end{displaymath}

It is easy to see that the result in equation (\ref{e-loss-stable-ave}) is real. Since the electric field and electric induction are both real in coordinate space, it follows from Eq.~(\ref{D-vs-E-x}) that the dielectric tensor obeys the relations 
\be
\label{sym-rel}
\Re \varepsilon^{ij}(- \omega,-{\bf k}) = \Re \varepsilon^{ij}(\omega,{\bf k}) ,
~~~~~~~~~~~~
\Im \varepsilon^{ij}(- \omega,-{\bf k}) = - \Im \varepsilon^{ij}(\omega,{\bf k}).
\ee
Since the analogous relations hold for the matrix (propagator) $\Sigma^{-1}(\omega,{\bf k})$, the real and imaginary contributions to the integrand in Eq.~(\ref{e-loss-stable-ave}) are, respectively, odd and even as functions of ${\bf k}$. Therefore, only the imaginary part of  $\Sigma^{-1}(\omega,{\bf k})$, which is responsible for dissipative phenomena, contributes to the integral (\ref{e-loss-stable-ave}), and the energy loss is real as it should be. 

In an isotropic plasma the dielectric tensor can be decomposed into longitudinal and transverse components
\be
\varepsilon^{ij}(\omega,{\bf k}) = 
\varepsilon_L(\omega,{\bf k}) \: \frac{k^ik^j}{{\bf k}^2}
+ \varepsilon_T(\omega,{\bf k}) \:
\Big(\delta^{ij} - \frac{k^ik^j}{{\bf k}^2}\Big) ,
\ee
and the matrix $\Sigma^{ij}(\omega,{\bf k})$ can be inverted to obtain the propagator as
\be
\label{inv-sigma-iso}
(\Sigma^{-1})^{ij}(\omega,{\bf k}) = 
\frac{1}{\omega^2 \varepsilon_L(\omega,{\bf k})}
\frac{k^ik^j}{{\bf k}^2}
+ \frac{1}{\omega^2 \varepsilon_T(\omega,{\bf k})-{\bf k}^2}
\Big(\delta^{ij} - \frac{k^ik^j}{{\bf k}^2}\Big) .
\ee
Substituting this expression into Eq.~(\ref{e-loss-stable-ave}), the energy loss is written
\ba
\label{e-loss-isotropic}
\frac{d \overline{E}}{dt} = -i g^2 C_R 
\int {d^3k \over (2\pi)^3} \;
\frac{\bar{\omega}}{{\bf k}^2} \;
\bigg[
\frac{1}
{\varepsilon_L(\bar{\omega},{\bf k})}
+ \frac{{\bf k}^2{\bf v}^2 - \bar{\omega}^2}
{\bar{\omega}^2
\varepsilon_T(\bar{\omega},{\bf k})-{\bf k}^2}
\bigg] .
\ea
Using the symmetry relations (\ref{sym-rel}) for $\varepsilon_{L,T}(\omega,{\bf k})$, Eq.~(\ref{e-loss-stable-ave}) becomes
\ba
\label{e-loss-isotropic-final}
\frac{d \overline{E}}{dt} = - g^2 C_R 
\int {d^3k \over (2\pi)^3} \;\frac{\bar{\omega}}{{\bf k}^2} \;
\bigg[
\frac{{\rm Im}\, \varepsilon_L(\bar{\omega},{\bf k})}
{|\varepsilon_L(\bar{\omega},{\bf k})|^2}
+ \frac{\bar{\omega}^2( {\bf k}^2{\bf v}^2 - \bar{\omega}^2) {\rm Im}\, \varepsilon_T(\bar{\omega},{\bf k})}
{|\bar{\omega}^2 \varepsilon_T(\bar{\omega},{\bf k})-{\bf k}^2|^2} 
\bigg] .
\ea

As discussed under equation (\ref{e-loss-2}), the energy loss formula (\ref{e-loss-2}), and consequently the formula (\ref{e-loss-main}), includes the self-interaction of the test parton with the electric field generated by the parton's current (\ref{current-x}). The parton's self-interaction should not contribute to the energy loss (\ref{e-loss-isotropic}), and therefore we need to calculate this contribution separately and, if it is not zero, we need to subtract it from the energy loss obtained from Eq.~(\ref{e-loss-isotropic}).  Since the effect of self-interaction is the same in a vacuum and in a medium, we derive it substituting into Eq.~(\ref{e-loss-isotropic}) the dielectric functions of the vacuum, which are
\ba 
\label{e-vacuum}
\varepsilon_L(\omega,{\bf k}) = \varepsilon_T(\omega,{\bf k})=1 .
\ea 
Using Eq.~(\ref{e-vacuum}) the formula (\ref{e-loss-isotropic}) gives
\ba
\nn
\frac{d\overline{E}}{dt}\bigg|_{\rm vacuum} &=&
i g^2 C_R (1 - v^2)
\int {d^3k \over (2\pi)^3} \;
\frac{\bar{\omega} }
{\bar{\omega}^2 -{\bf k}^2}
\\ [2mm]\label{e-loss-isotropic-vacuum-2}
&= &
-i { g^2 C_R \over (2\pi)^2} (1-v^2)
\int_0^\infty dk\, k \int_{-1}^{+1} \frac{d(\cos \theta) \,\cos \theta}{1-v^2 \cos^2 \theta} = 0,
\ea
where we have chosen the axis $z$ along the vector ${\bf v}$ and written $\bar \omega = {\bf k}\cdot{\bf v} = k \cos\theta$. Although the momentum integral is quadratically divergent, the angular integral vanishes and the three-dimensional integral is zero. The zero result is also expected from Eq.~(\ref{e-loss-isotropic-final}) because the vacuum dielectric functions (\ref{e-vacuum}) are purely real. Thus we see that the parton's self-interaction does not contribute to the equilibrium energy loss formula (\ref{e-loss-isotropic}) or (\ref{e-loss-isotropic-final}). In Sec.~\ref{sec-vacuum} we will show that this is not the case when the energy loss calculation is formulated as an initial value problem. 

The result (\ref{e-loss-isotropic}) or (\ref{e-loss-isotropic-final}) agrees with the expression obtained in \cite{Mrowczynski:1991da} using kinetic theory, and with the result for the energy loss due to soft collisions calculated in the HTL approximation \cite{Braaten:1991we}, see also the textbook \cite{lebellac}. However, this is not the complete energy loss but rather the soft contribution to it when the wave vector ${\bf k}$ is of the order of the Debye mass. Physically it corresponds to an interaction of the test parton with soft collective excitations of the plasma medium. The incompleteness of the formula (\ref{e-loss-isotropic}) or (\ref{e-loss-isotropic-final}) is signaled by the logarithmic divergence as  $|{\bf k}| \rightarrow \infty$. To obtain the complete collisional energy loss, the formula (\ref{e-loss-isotropic}) should be combined with the hard contribution describing elastic collisions of the test parton with plasma constituents with momentum transfer much exceeding the Debye mass. The hard contribution is not ultraviolet divergent, as the maximal momentum transfer is constrained by the collision kinematics. The soft contribution to the energy loss depends logarithmically on the upper cut-off $k_{\rm max}$ divided by the Debye mass $m$, while the hard contribution has a logarithmic dependence on the energy of the parton $E$ divided by the same cut-off $k_{\rm max}$. The energy loss thus equals
\be
\frac{d\overline{E}}{dt} = X \ln\bigg(\frac{k_{\rm max}}{m}\bigg) + Y  \ln\bigg(\frac{E}{k_{\rm max}}\bigg) .
\ee
It can be shown \cite{Braaten:1991we,lebellac} that the coefficients $X,Y$ are equal to each other and therefore
\be
\frac{d\overline{E}}{dt} = X \ln\bigg(\frac{E}{m}\bigg) .
\ee
The result is that the cut-offs cancel and one obtains a good approximation to the energy loss from the soft contribution with the parton energy used as an upper cut-off.

\begin{figure}[t]
\centering
\includegraphics[width=0.60\textwidth]{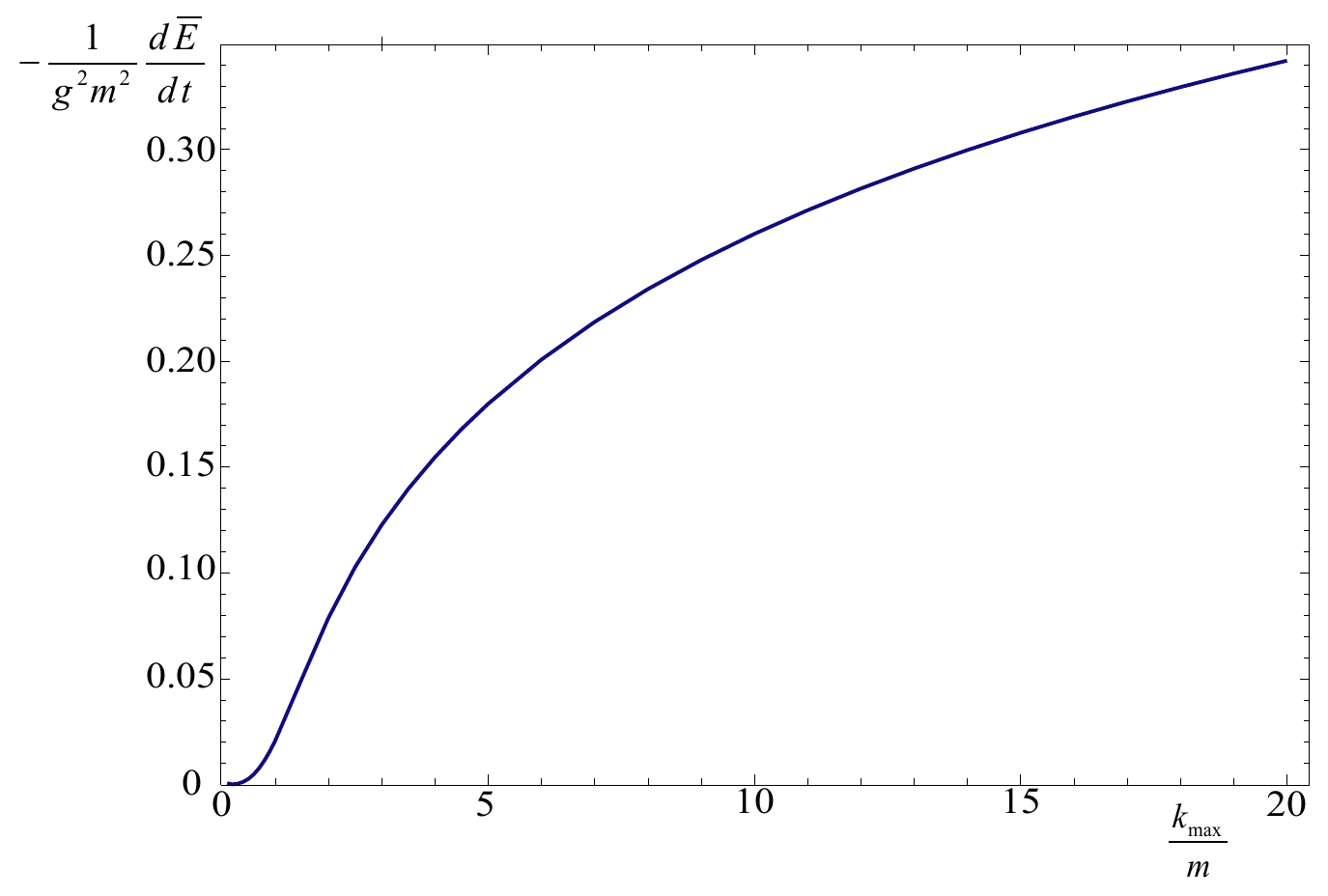}
\vspace{-5mm}
\caption{(Color online) The parton energy loss per unit time in equilibrium plasma as a function of $k_{\rm max}$.}
\label{fig-eq}
\end{figure}

As was mentioned in the introduction, the energy loss in anisotropic QGP was computed previously by Romatschke and Strickland \cite{Romatschke:2004au}.  Their result can be obtained from our formula (\ref{e-loss-stable-ave}) by using an anisotropic propagator for $\Sigma^{-1}(\omega,{\bf k})$ and including only the contribution from the pole $\omega = \bar{\omega}$. Clearly this procedure produces a result for the energy loss that is completely time independent. As we will see in the subsequent sections, the energy loss in anisotropic plasma is actually strongly time dependent because of the unstable modes. 

In order to compare our results for the energy loss in an unstable plasma to the corresponding equilibrium result, we have computed numerically the integral (\ref{e-loss-isotropic}) in spherical coordinates. As already mentioned, the integral is logarithmically divergent at large $k\equiv |{\bf k}|$, so we introduce a cutoff $k \le k_{\rm max}$. When studying plasmas with massless constituents, the Debye mass $m$ can be chosen as the only dimensionful parameter, and we therefore use a system of units where all dimensionful quantities are rescaled by the appropriate powers of $m$. Numerically one can simply set $m=1$. We define the Debye mass as
\be
\label{Debye-mass}
m^2 \equiv g^2 \int {d^3p \over (2\pi)^3} \, \frac{f({\bf p})}{|{\bf p}|}\,.
\ee
For isotropic momentum distributions, the formula  (\ref{Debye-mass}) coincides with the standard definition of the Debye mass. 
The same definition can also be used for the anisotropic plasmas studied in this paper. In Fig.~\ref{fig-eq} we show the energy loss in isotropic QGP divided by $g^2 m^2$ as a function of $\frac{k_{\rm max}}{m}$ computed for $C_R = N_c = 3$ which corresponds to a gluon. Since the energy loss is divided by $g^2 m^2$ we do not need to specify the value of $g$. The numbers from this figure will serve as a reference for our results on the energy loss in unstable plasmas.

\section{Initial conditions}
\label{sec-ini-cond}

When the plasma is anisotropic, the propagator $\Delta (\omega,{\bf k})=\Sigma^{-1}(\omega,{\bf k})$ has poles in the upper half-plane of complex $\omega$ which correspond to instabilities,  and the contributions to the energy loss from these poles grow exponentially in time. This means that the terms in Eq.~(\ref{e-loss-main}) which contain the initial values of fields ${\bf D}_0$ and ${\bf B}_0$ are amplified by an exponential factor and, in contrast to the equilibrium situation, they cannot, in general, be neglected. 

\subsection{Uncorrelated initial conditions}
\label{sec-uncorr-ini}

The simplest choice of the initial condition is ${\bf D}_0 = {\bf B}_0 = 0$, which means that the energy loss formula (\ref{e-loss-2}) becomes   
\ba
\label{e-loss-uncorr}
\frac{d\overline{E}(t)}{dt} = i g^2 C_R  v^i v^j 
\int {d^3k \over (2\pi)^3}
\int_{-\infty +i\sigma}^{\infty +i\sigma}
{d\omega \over 2\pi i}
e^{-i(\omega - \bar{\omega}) t} \frac{\omega}{\omega - \bar{\omega}} \,
(\Sigma^{-1})^{ij}(\omega,{\bf k}) ,
\ea
where we have used the relation (\ref{ave-color-charge-2}) to average over colors. In fact, the formula (\ref{e-loss-uncorr}) holds for a whole class of initial conditions whenever ${\bf D}_0$ and ${\bf B}_0$ are {\em independent} of the test parton's current. In this case, the contributions to the energy loss (\ref{e-loss-main}) which contain ${\bf D}_0$ and ${\bf B}_0$ are linear in the parton's color charge $Q_a$,  and consequently they vanish when color averaging is performed using  the relation (\ref{ave-color-charge-1}). 

Physically this result can be understood as follows. Let us consider an electron moving in an external electromagnetic field which is independent of the current generated by the electron. The energy loss formula is given by the  electromagnetic analog of the formula (\ref{e-loss-1}) where ${\bf E}\big(t,{\bf r}(t)\big)$ is the external electric field along the electron's trajectory.  The electromagnetic analog of averaging over the parton's color is the averaging over the possible charges of a hypothetical electron which could carry either negative or positive charge, or the averaging over the charges of an electron and a positron. If an electron's energy increases by $\Delta E$ in the time interval  $\Delta t$, a positron's energy would decreases by $-\Delta E$ in the same field configuration and time interval. Therefore, after averaging over charges, the net change in the energy is zero. 

It is important to remember that the contribution to the energy loss from the first term in (\ref{e-loss-main}), which is proportional to the current and not the initial fields, is non zero, even when uncorrelated initial conditions are used. Mathematically, this happens because this term is proportional to the square of the charge. In an electromagnetic plasma, $e^2$ is strictly positive, and in a QCD plasma the factor 
 $Q^a Q^a$ does not give zero when averaged (see equation (\ref{ave-color-charge-2})). Physically we see that the energy losses of the electron and positron have same sign because they are not interacting with external fields which are independent of their currents, but instead with the electric fields which they have induced in the medium. We also note that the procedure of averaging over electric charges looks similar to that of averaging over colors but the physical situation is quite different. A color charge is gauge dependent and consequently it is not a physical observable. Therefore, the averaging over colors must be performed in order for the energy loss to have a physical meaning.

It is interesting to note that we can obtain the same energy loss formula (\ref{e-loss-uncorr}) in a different way. If we multiply the current in equation (\ref{current-x}) by a step function $\Theta(t)$ and then repeat the whole calculation using the usual two-sided Fourier transformation, the identical result is found. The initial fields ${\bf D}_0$ and ${\bf B}_0$ do not appear in the two-sided Fourier transformed Maxwell equations, and the two-sided Fourier transform of the current with the additional step function is the same as the one-sided Fourier transform of original current. Although the same result can be obtained in two different ways, the physical interpretation of the two procedures is somewhat different.  Using the two-sided transformation with the current multiplied by a step function, we assume that the plasma system exists for all times but the test parton appears in the plasma at $t=0$. This was the problem studied in the papers \cite{Peigne:2005rk,Adil:2006ei}. When the one-sided Fourier transformation is used, it is understood that we observe the whole system, which includes the plasma and the test parton, starting only at $t=0$. The initial values of the fields ${\bf D}_0$ and ${\bf B}_0 $ can be chosen to be independent of the parton's color state, but they could also be specified differently. In the next section we consider a class of nontrivial initial conditions for which the fields ${\bf D}_0$ and ${\bf B}_0$ are strongly correlated with the current generated by the test parton.

\subsection{Correlated initial conditions}
\label{sec-corr-ini}

We have shown in the previous section that if the initial conditions are chosen in any way that is independent of the parton's current, they will not contribute to the energy loss. In this section we will consider another kind of initial conditions.  First we note that although initial conditions are always required to solve differential equations, they are usually determined by physical arguments which go beyond the differential equations under consideration. We argue below that in the energy loss calculation, a kind of correlated initial conditions might be the most physical. We imagine that a process which is responsible for the occurrence of a test parton in a plasma system at the time $t=0$, also polarizes the medium producing a chromodynamic field which is then correlated with the parton's color state. We would like to see if the energy loss is sensitive to this kind of {\it correlated} initial condition.

It is important to realize that the history of the system under consideration does not start at $t=0$ when the test parton enters the plasma but {\em earlier} when the colliding nuclei begin to approach each other. When they collide, they produce among other particles the test parton. The plasma is produced at more or less the same time as the test parton, and therefore, there is no reason to assume that the test parton and plasma, which are both a part of a bigger system, are completely uncorrelated. Note that this correlation is generated before the collision at $t=0$, and therefore there is no violation of causality.

Although the physical origin of a correlation between the parton and the plasma fields is clear, the question of how to express this correlation mathematically is much more difficult. To derive an upper limit, we assume that the parton enters the system in the remote past at $t= -\infty$, observing that the parton's current (\ref{current-x}) can be extended to the time interval from $-\infty$ to $\infty$. Flying across the plasma, the parton polarizes the medium and induces a chromodynamic field. The initial fields ${\bf D}_0$ and ${\bf B}_0$ are identified with the induced fields at $t=0$.

To determine the fields ${\bf D}_0$ and ${\bf B}_0$ we solve the Maxwell equations (\ref{Maxwell-eqs-x-lin-1}) and (\ref{Maxwell-eqs-x-lin-2}) using a normal (two-sided) Fourier transform with the time integral from $-\infty$ to $\infty$. We use tildes to indicate that a two-sided Fourier transform was taken, which means that, for example, $D(\omega,{\bf k})$ and $\tilde D(\omega,{\bf k})$ are different functions of $\omega$ but the same function of ${\bf k}$.  However, we note that $\tilde\varepsilon(\omega,{\bf k})=\varepsilon(\omega,{\bf k})$ and $\tilde\Sigma^{-1}(\omega,{\bf k})=\Sigma^{-1}(\omega,{\bf k})$ because these functions obey the retarded initial condition and therefore $\varepsilon(t,{\bf r}) = \Sigma^{-1}(t,{\bf r}) = 0$ for $t < 0$.
Solving the equations  (\ref{Maxwell-eqs-x-lin-1}) and (\ref{Maxwell-eqs-x-lin-2}) using a two-sided Fourier transform produces the result in equation (\ref{E-field-k})  with $E^i_a$ and $j^j_a$ tilded and $B^l_{0a}=D^j_{0a}=0$,
since the initial fields in Eq.~(\ref{E-field-k}) come from the $t=0$ lower limit in the one-sided Fourier transform. Using the tilded version of the material relation (\ref{D-vs-E}), the electric induction is
\ba 
\label{D-tilde}
\tilde D_a^i(\omega,{\bf k}) = -i  \,
 \omega \, \varepsilon^{ij}(\omega,{\bf k})
(\Sigma^{-1})^{jk}(\omega,{\bf k})
\tilde j_a^k(\omega,{\bf k}) ,
\ea
where the two-sided Fourier transform of the current in equation (\ref{current-x}) is
\be
\tilde {\bf j}_a(\omega,{\bf k}) = g Q^a {\bf v} 2\pi \delta (\omega - \bar{\omega}) .
\ee 
Taking the inverse two-sided Fourier transform of the result (\ref{D-tilde}), we obtain
\ba 
 D_a^i(t,{\bf k}) = \int^\infty_{-\infty} \frac{d\omega}{2\pi}\,e^{-i\omega t}
\tilde D_a^i(\omega,{\bf k}) = -i e^{-i\bar\omega t}g Q^a \bar{\omega} \,
 \varepsilon^{ij}(\bar{\omega},{\bf k})
(\Sigma^{-1})^{jk}(\bar{\omega},{\bf k})  v^k ,
\ea
and setting $t=0$, we arrive at
\be
\label{D_0-2}
D^i_{0a}({\bf k}) =-i g Q^a \bar{\omega} \,
 \varepsilon^{ij}(\bar{\omega},{\bf k})
(\Sigma^{-1})^{jk}(\bar{\omega},{\bf k})  v^k .
\ee
Using the same method we obtain the initial value of the chromomagnetic field 
\be
\label{B_0}
B^i_{0a}({\bf k}) 
= -i g Q^a  \epsilon^{ijk}k^j
(\Sigma^{-1})^{kl}(\bar{\omega},{\bf k})  v^l .
\ee

The formulas (\ref{D_0-2}) and (\ref{B_0}) provide maximally correlated initial conditions.  In order to consider initial conditions with differing degrees of correlation, we will multiply the initial fields (\ref{D_0-2}) and (\ref{B_0}) by a phase factor $\cos\alpha \in [-1,1]$. The choices $\cos\alpha =\pm 1$ correspond to maximally correlated and anticorrelated initial fields. These two extreme cases provide limits on the possible effects of correlated initial conditions. 
We substitute the initial fields ${\bf D}_0$ and ${\bf B}_0$ given by Eqs.~(\ref{D_0-2}) and (\ref{B_0}) into the energy loss formula (\ref{e-loss-main}) and insert the phase factor $\cos\alpha$ as described above. After averaging over the parton's color we obtain
\ba
\label{e-loss-corr}
\frac{d\overline{E}(t)}{dt} &=& i g^2 C_R  v^i v^l 
\int {d^3k \over (2\pi)^3}
\int_{-\infty +i\sigma}^{\infty +i\sigma}
{d\omega \over 2\pi i}
e^{-i(\omega - \bar{\omega}) t}
(\Sigma^{-1})^{ij}(\omega,{\bf k})
\\ \nn
&\times&
\bigg\{
\frac{\omega \delta^{jl}}{\omega - \bar{\omega}}
- \cos\alpha \Big[  (k^j k^k - {\bf k}^2 \delta^{jk})
(\Sigma^{-1})^{kl}(\bar{\omega},{\bf k}) 
- \omega \, \bar{\omega} \, \varepsilon^{jk}(\bar{\omega},{\bf k})
(\Sigma^{-1})^{kl}(\bar{\omega},{\bf k})   
 \Big] \bigg\} \,.
\ea
This result, which reduces to the formula (\ref{e-loss-uncorr}) when $\cos\alpha =0$, will be further studied in the subsequent sections for two different unstable plasma systems. 

\section{Self-interactions}
\label{sec-vacuum}

As already discussed in detail in the context of the equilibrium result (\ref{e-loss-isotropic-final}), the energy loss formulas include the effect of self-interaction -- also called the vacuum effect -- which needs to be subtracted if it is non-zero. In this section we calculate the self-interaction contribution to the energy loss given by Eq.~(\ref{e-loss-uncorr}) and (\ref{e-loss-corr}). We follow the same method as in Sec.~\ref{sec-stable}. We evaluate the formulas (\ref{e-loss-uncorr}) and (\ref{e-loss-corr}) with the propagator $\Sigma^{-1}(\omega,{\bf k})$ in the form (\ref{inv-sigma-iso}) with the vacuum dielectric functions (\ref{e-vacuum}).  However, the calculation is not the same as the one done in Sec.~\ref{sec-stable}. The equilibrium result (\ref{e-loss-isotropic-final}) only has a contribution from the pole $\omega = \bar\omega$, but the energy loss formulas (\ref{e-loss-uncorr}) and (\ref{e-loss-corr}) with a vacuum propagator also include contributions from the poles $\omega = 0$ and $\omega = \pm |{\bf k}|$, which make the effect of self-interaction time dependent.
We discuss only the vacuum contribution to the energy loss formula for correlated initial conditions (\ref{e-loss-corr}), because the corresponding result for uncorrelated initial conditions can be obtained by setting $\cos\alpha =0$.

To compute the vacuum effect we substitute into the formula (\ref{e-loss-corr}) the vacuum propagator (\ref{inv-sigma-iso}) with $\varepsilon_L(\omega,{\bf k}) = \varepsilon_T(\omega,{\bf k}) =1$. In this way one finds the longitudinal part 
\ba
\label{e-loss-vac-L1}
\frac{d\overline{E}_L(t)}{dt}\bigg|_{\rm vacuum} =  i g^2 C_R 
\int {d^3k \over (2\pi)^3} 
\frac{\bar{\omega}}{k^2} 
\int_{-\infty +i\sigma}^{\infty +i\sigma}
{d\omega \over 2\pi i}
\frac{e^{-i(\omega - \bar{\omega}) t}}{\omega}
\Big[ \frac{\bar\omega}{\omega -\bar\omega} 
+  \cos\alpha  \Big] ,
\ea
and the transverse one
\ba
\label{e-loss-vac-T1}
\frac{d\overline{E}_T(t)}{dt}\bigg|_{\rm vacuum} =  
 i g^2 C_R 
\int {d^3k \over (2\pi)^3} 
\Big(1 - \frac{\bar\omega^2}{k^2}  \Big) 
\int_{-\infty +i\sigma}^{\infty +i\sigma}
{d\omega \over 2\pi i}
\frac{e^{-i(\omega - \bar{\omega}) t}}{\omega^2 - k^2} 
\bigg[
\frac{\omega}{\omega - \bar{\omega}} 
+ \cos\alpha \,
\frac{\omega \bar{\omega} + k^2}{\bar{\omega}^2 - k^2} \bigg],
\ea
where $k \equiv  |{\bf k}|$.
Performing the integral over $\omega$, which includes contributions from the poles at $\omega = \bar\omega$ and $\omega = 0$ in case of the longitudinal part (\ref{e-loss-vac-L1}), and the poles at $\omega = \bar{\omega}$  and $\omega = \pm k$ in case of the transverse one (\ref{e-loss-vac-T1}), we obtain
\ba
\label{e-loss-vac-L2}
\frac{d\overline{E}_L(t)}{dt} \bigg|_{\rm vacuum}
&=& -  (1-\cos\alpha) g^2 C_R
\int {d^3k \over (2\pi)^3}
\frac{\bar{\omega} \,\sin(\bar{\omega}t)}{k^2} ,
\\[2mm]
\label{e-loss-vac-T2}
\frac{d\overline{E}_T(t)}{dt}\bigg|_{\rm vacuum} 
&=& - (1- \cos\alpha) \frac{i g^2 C_R}{2}
\int {d^3k \over (2\pi)^3} 
\Big(1 - \frac{\bar{\omega}^2}{k^2}  \Big) \Big(
\frac{e^{-i(k - \bar{\omega}) t}}{k - \bar{\omega}} 
- \frac{e^{i(k + \bar{\omega}) t}}{k + \bar{\omega}} 
\Big) .
\ea
We note that the pole $\omega = \bar{\omega}$  does not actually contribute to the transverse part (\ref{e-loss-vac-T2}) because the integrand is odd as a function of ${\bf k}$ and therefore it gives zero when integrated.

The integrals over ${\bf k}$ are calculated in spherical coordinates using an upper cut-off $k_{\rm max}$. Summing the longitudinal and transverse parts, the complete vacuum contribution to the energy loss formula (\ref{e-loss-corr}) equals
\ba
\label{e-loss-vac-final}
\frac{d\overline{E}(t)}{dt}\bigg|_{\rm vacuum} = -\frac{(1-\cos\alpha) g^2 C_R}{4\pi^2 t^2}
\Big[2\big( {\rm Si}(k_{\rm max} t) -\sin (k_{\rm max} t)\big) 
+ \big(2k_{\rm max} t-{\rm Si}(2k_{\rm max} t) \big)\Big],
\ea
where ${\rm Si}(z)$ is the sine integral defined as 
\be
{\rm Si}(z) \equiv \int^z_0dx \frac{\sin(x)}{x} .
\ee 
The first term in the expression (\ref{e-loss-vac-final}) is the longitudinal part and the second term represents the transverse piece which linearly diverges with increasing $k_{\rm max}$. Both the longitudinal and transverse contributions go to zero when $t \rightarrow 0$. The vacuum contribution to the energy loss formula with uncorrelated initial conditions (\ref{e-loss-uncorr}) is given by Eq.~(\ref{e-loss-vac-final}) with  $\cos\alpha=0$. From Eq.~(\ref{e-loss-vac-final}) it is clear that the vacuum contribution is not zero unless we choose maximally correlated initial conditions (for which  $\cos\alpha=1$), and therefore the self-interaction effect must be subtracted from the energy loss formula in all other cases.

\section{Anisotropic plasmas}
\label{sec-aniso-plamsas}

The energy loss in isotropic plasmas has been calculated from the general formula (\ref{e-loss-main}) in section \ref{sec-stable}. In this section we develop our formalism to apply it to a general class of anisotropic momentum distributions of plasma constituents which was introduced in \cite{Romatschke:2003ms} and has been used in various studies of QGP, see {\it e.g.} \cite{Romatschke:2004jh,Attems:2012js,Florkowski:2012as,Dumitru:2007hy,Martinez:2008di,Schenke:2006yp}. These anisotropic distributions are obtained from the isotropic one by deforming it -- squeezing or stretching -- in one direction. The dispersion relations of the collective modes, which are needed to compute the energy loss, have been studied in great detail in our recent study \cite{Carrington:2014bla} for all possible degrees of  deformation from the extremely prolate case, when the momentum distribution is infinitely elongated in one direction, to the extremely oblate distribution, which is infinitely squeezed in one direction. 

In our derivation of the energy loss formula, and our calculation of energy loss for isotropic systems, we have mostly used the terminology of classical electrodynamics of continuous media, with the dielectric tensor playing a key role. From now on we will switch to the language of quantum field theory, and make use of the polarization tensor and gluon propagator which were already introduced in Eqs.~(\ref{diel-tensor}) and (\ref{prop}). The two languages are equivalent, as QCD in the hard-loop approximation is essentially classical, but the terminology of quantum field theory is more commonly used when working with the distribution introduced in \cite{Romatschke:2003ms}.

In order to simplify the notation, in the rest of this section we omit the arguments which denote dependence on the wave vector. For example, we write $\alpha(\omega,{\bf k})$ as $\alpha(\omega)$, $\Delta^{ij}(\omega,{\bf k})$ as $\Delta^{ij}(\omega)$, {\it etc. }

\subsection{Propagator}

To compute the energy loss using the formula (\ref{e-loss-main}) we have to invert the matrix $\Sigma$ defined by Eq.~(\ref{matrix-sigma}) or (\ref{prop}), which is the inverse gluon propagator in the temporal axial gauge. In isotropic plasmas the matrix depends on only one vector ${\bf k}$. It can be decomposed into transverse and longitudinal components and is therefore easily inverted giving  Eq.~(\ref{inv-sigma-iso}). We will now consider momentum distributions of the plasma constituents that can be obtained from the isotropic one by deforming it in one direction along the unit anisotropy vector  ${\bf n}$. In this case the matrix $\Sigma$ depends on two vectors ${\bf k}$ and ${\bf n}$, and it is symmetric $\Sigma^{ij} = \Sigma^{ji}$. To invert such a matrix, we introduce, following \cite{Kobes:1990dc}, the vector ${\bf n}_T$  
\be
\label{n-T}
n_T^i = \big(\delta^{ij} - \frac{k^i k^j}{{\bf k}^2}\big) \, n^j
\ee 
and we define four tensors \cite{Romatschke:2003ms}
\ban
\label{decomp}
A^{ij} &=& 
\delta^{ij} - \frac{k^i k^j}{{\bf k}^2},
\;\;\;\;\;\;
B^{ij} = 
\frac{k^i k^j}{{\bf k}^2} ,
\\[2mm] 
C^{ij} &=& 
\frac{n_T^i n_T^j}{{\bf n}_T^2},
\;\;\;\;\;\;\;\;\;
D^{ij} = 
k^i n_T^j + k^j n_T^i\,.
\ean
The matrix $\Sigma$ defined in Eq.~(\ref{prop}) is written
\be
\label{Sigma-A-B-C-D}
\Sigma^{ij}(\omega) = \big(\omega^2-{\bf k}^2-\alpha(\omega)\big) A^{ij}
+ \big(\omega^2-\beta(\omega)\big) B^{ij} 
- \gamma(\omega) C^{ij} 
- \delta(\omega) D^{ij} ,
\ee
where $\alpha, \; \beta, \; \gamma, \; \delta$ are the coefficients of the decomposition of the polarization tensor
\be
\label{pi-decomposition}
\Pi^{ij}(\omega) = \alpha(\omega) A^{ij}+\beta(\omega) B^{ij} + \gamma(\omega) C^{ij} + \delta(\omega) D^{ij} .
\ee
Writing the propagator $\Delta \equiv \Sigma^{-1}$ in the same basis, the equation  $\Delta \Sigma =1$ gives
\ba 
\nn
\Delta^{ij} (\omega) &=& \Delta_A(\omega) \, (A^{ij}-C^{ij}) 
\\[2mm]
\label{propagator-DA-DG}
&+&  \Delta_G(\omega) 
\Big(\big(\omega^2 - {\bf k}^2 -\alpha(\omega) - \gamma(\omega) \big)B^{ij} 
+ \big(\omega^2-\beta(\omega) \big) C^{ij} + \delta D^{ij} \Big),
\ea
where 
\ba
\label{Delta-A}
\Delta_A^{-1}(\omega) &\equiv &  \omega^2 - {\bf k}^2 - \alpha(\omega)  ,
\\[2mm]
\label{Delta-G}
\Delta^{-1}_G(\omega) &\equiv& \big(\omega^2 - \beta(\omega)\big)
\big(\omega^2 - {\bf k}^2 - \alpha(\omega) - \gamma(\omega)\big) 
- {\bf k}^2  {\bf n}_T^2 \delta^2(\omega).
\ea
From Eq.~(\ref{propagator-DA-DG}) we see that the poles of the propagator, which correspond to gluon collective modes, or plasmons, are given by the dispersion equations 
\be
\label{dis-eq-general-2}
\Delta_A^{-1}(\omega) =0, ~~~~~~~\Delta_G^{-1}(\omega) =0 .
\ee
The equations (\ref{dis-eq-general-2}) are obviously equivalent  to the general dispersion equation (\ref{dis-eq-general}).

\subsection{Integrand}
\label{sec-integrand}

Substituting the propagator (\ref{propagator-DA-DG}) into the energy loss formula (\ref{e-loss-corr}) and contracting all indices, we obtain an expression that we will use to do calculations for the extremely prolate and extremely oblate momentum distributions discussed in Sec.~\ref{sec-ex-pro-obla}. We use a spherical coordinate system with the $z$-axis along the anisotropy vector ${\bf n}$. The angles $\theta$ and $\phi$ are the zenithal and azimuthal angles of the vector ${\bf k}$, and $\Theta$ is the angle between the velocity of the test parton ${\bf v}$ and the anisotropy vector ${\bf n}$. In our coordinate system the vectors ${\bf n}$, ${\bf v}$ and ${\bf k}$ are
\ba
\nn
{\bf n} &=& (0,0,1) ,
\\ 
\label{coordinate-system}
{\bf v} &=& (\sin\Theta,0,\cos\Theta) ,
\\ \nn
{\bf k} &=& k (\sin\theta \cos\phi,\sin\theta \sin\phi, \cos\theta) .
\ea
The energy loss formula (\ref{e-loss-corr}) is written as 
\ba
\label{e-loss-Integrand}
\frac{d\overline{E(t)}}{dt} = 
g^2C_R \int {d^3k \over (2\pi)^3} 
\int^{\infty+i\sigma}_{-\infty+i\sigma}\frac{d\omega}{2\pi i} \;e^{-it(\omega-\bar\omega)}\; {\rm Integrand}\, ,
\ea
where the integrand is divided into several different pieces by writing it as
\ba
{\rm Integrand} = A_j+G_j+ \cos\alpha \big( A_{\rm{ic}}+G_{\rm{ic}}  +[AA]_{\rm{ic}}+[GG]_{\rm{ic}} \big) .
\ea
The two terms $A_j$ and $G_j$ are the contributions from the first term in the square bracket in Eq.~(\ref{e-loss-main}), which comes from the parton current (\ref{current-k}). When the initial fields are set to zero, or when we have uncorrelated initial conditions, these are the only terms that survive. They are proportional to $\Delta^{-1}_A(\omega)$ and $\Delta^{-1}_G(\omega)$. The terms $A_{\rm ic}$, $G_{\rm ic}$, $[AA]_{\rm ic}$, $[GG]_{\rm ic}$ are the contributions from the second two terms in the square bracket in Eq.~(\ref{e-loss-main}) and come from the initial fields. They are proportional to $\Delta_A(\omega)$, $\Delta_G(\omega)$, $\Delta_A(\omega) \,\Delta_A(\bar\omega)$ and $\Delta_G(\omega)\, \Delta_G(\bar\omega)$.  In the future we will refer to the first two terms in the formula (\ref{e-loss-Integrand}) as `current contributions,' and the last four terms will be called `field contributions.'
After performing all contractions we obtain
\ban
 A_j &\equiv& \frac{ \hat\omega \big[ \tilde\omega^2 (x^2-1)-x^2-Y^2+1\big]}
{(1-x^2) (\hat\omega - \tilde\omega) \Delta^{-1}_A(\omega)} ,
\\[2mm]
G_j &\equiv& \hat\omega \, \frac{Y^2 \beta^\prime(\omega)-\tilde\omega
 (x^2-1) \big\{ \tilde\omega \big[ k^2 (\hat\omega^2-1) - \alpha (\omega) - \gamma (\omega)\big] 
+ 2 k Y \delta (\omega)\big\}}
{\left(1-x^2\right) (\hat\omega-\hat\omega) \Delta^{-1}_G(\omega)} ,
\\[2mm]      
A_{\rm ic} &\equiv& \Big(\frac{\hat\omega}{\tilde\omega} - 1\Big) A_j ,
\\[2mm]
G_{\rm ic} &\equiv& \Big(\frac{\hat\omega}{\tilde\omega} - 1\Big) G_j ,
\\[2mm]
[AA]_{\rm ic} &\equiv& \frac{k^2 (\hat\omega-\tilde\omega)(\hat\omega+\tilde\omega)}
{\hat\omega \tilde\omega \Delta^{-1}_A(\bar\omega)} A_j ,
\\[2mm]
[GG]_{\rm ic} &\equiv& k^2 (\hat\omega+\tilde\omega) \,
\frac{k Y \tilde\omega \left(1-x^2\right)
\big[ \beta^\prime(\bar\omega) \delta (\omega)+\beta^\prime(\omega) \delta (\bar\omega) \big]
+ Y^2 \beta^\prime(\omega) \beta^\prime(\bar\omega)
+ k^2 \tilde\omega^2 \left(1-x^2\right)^2 \delta(\omega) \delta (\bar\omega)}
{ \tilde\omega \left(1-x^2\right) \Delta^{-1}_G(\omega) \Delta^{-1}_G(\bar\omega)} ,
\ean
where we have used the symbols
$$
x \equiv \cos\theta, ~~~~~~~~ \hat\omega \equiv \omega/k,
 ~~~~~~~~ \tilde\omega \equiv \bar\omega/k,  ~~~~~~~~ Y \equiv\cos\Theta - x \tilde\omega\,,
$$
and defined the function $\beta^\prime(\omega) \equiv \omega^2 - \beta(\omega)$.

\section{Extremely prolate and oblate plasmas}
\label{sec-ex-pro-obla}

In the early stages of a heavy ion collision, when partons are initially released from the incoming nucleons, the momentum distribution is strongly elongated along the beam - it has a {\em prolate} shape with the average transverse momentum much smaller than the average longitudinal one. Due to free streaming, see {\it e.g.} \cite{Jas:2007rw}, the distribution evolves in the local rest frame to a form which is squeezed along the beam - it has {\em oblate} shape with the average transverse momentum being much larger than the average longitudinal one. Prolate and oblate distributions can be obtained from an isotropic one by stretching or squeezing in the direction of a unit vector ${\bf n}$ which is chosen parallel to the beam direction.  In this paper we consider an {\em extremely} prolate and an {\em extremely} oblate distribution which are defined as
\ba
\label{extreme-prolate} 
f_{\rm ex-prolate}({\bf p}) &=& \delta(p_T)\, \frac{|p_L|}{p_T} \, g(p_L) ,
\\[2mm]
\label{extreme-oblate} 
f_{\rm ex-oblate}({\bf p}) &=& \delta(p_L) \, h (p_T), 
\ea
where $p_L \equiv {\bf p}\cdot{\bf n}$ and $p_T \equiv |{\bf p} - ({\bf p}\cdot{\bf n}){\bf n}|$. The functions $h(p_T)$ and $g(p_L)$ are determined from the normalization condition (\ref{Debye-mass}) which now has the form
\be
\label{mass-defn}
m^2 = {g^2 \over 4\pi^2} \int_0^\infty dp_T\, h(p_T) 
= {g^2 \over 4\pi^2} \int_{-\infty}^\infty dp_L \, g(p_L) .
\ee

\begin{figure}[t]
\center
\includegraphics[width=0.55\textwidth]{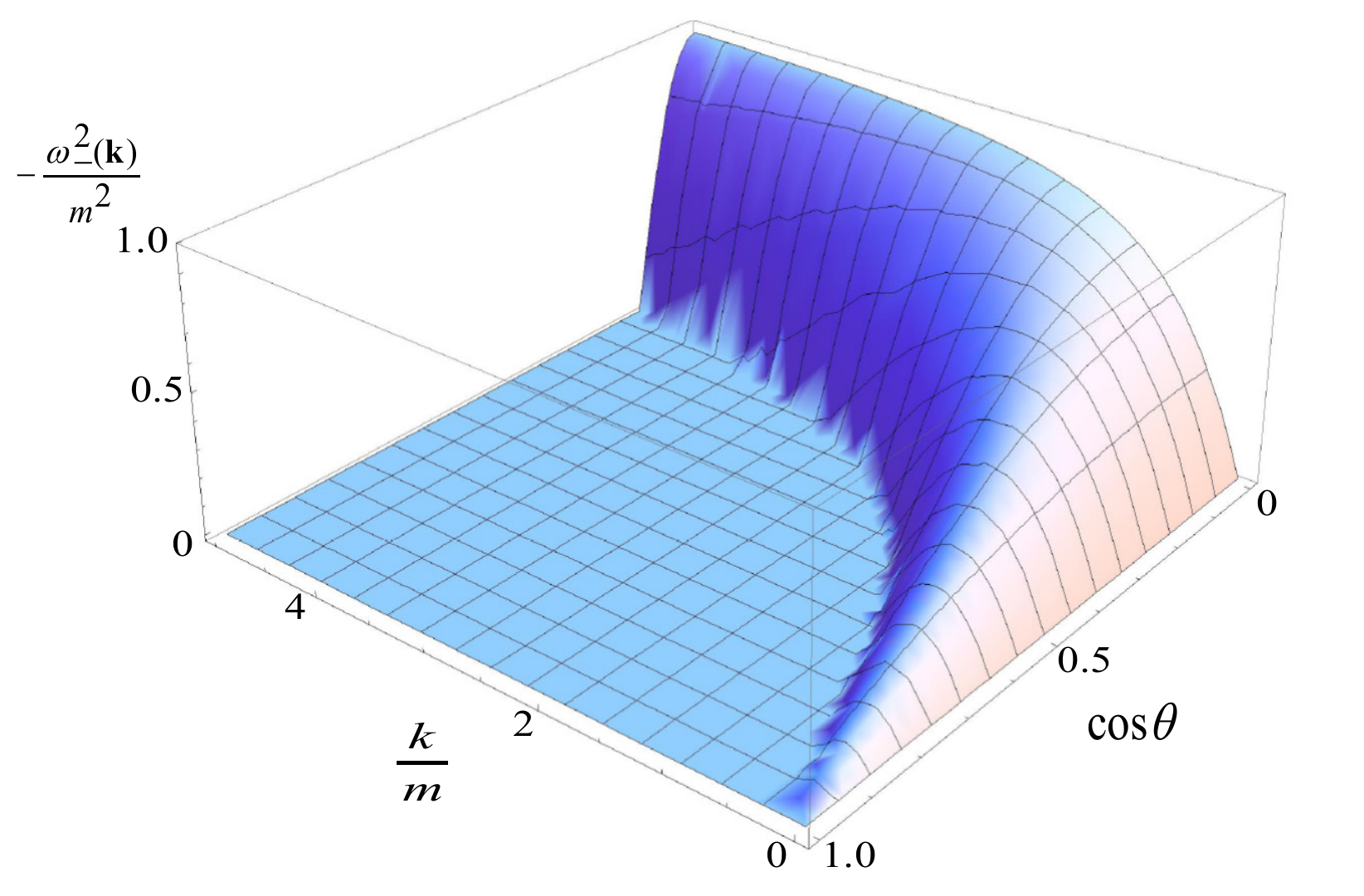}
\vspace{-3mm}
\caption{(Color online) Unstable mode for the extremely prolate plasma:  $-\omega_-^2({\bf k})$ as a function of $k$ and $\cos\theta$ in the domain  where $\omega_-^2({\bf k}) < 0$. The angle $\theta$ is between the vectors ${\bf k}$ and ${\bf n}$.}
\label{fig-omega-minus-ex-prolate}
\end{figure}

The collective modes are determined by substituting the distributions (\ref{extreme-prolate}) and (\ref{extreme-oblate})  into the dielectric tensor (\ref{eij}) which is related to the polarization tensor by Eq.~(\ref{diel-tensor}). One calculates the components of the polarization tensor (\ref{pi-decomposition}) and solves the dispersion equations (\ref{dis-eq-general-2}). The spectrum of plasmons of the extremely prolate and extremely oblate plasmas was analyzed in detail in our study \cite{Carrington:2014bla}. Below we give a summary of the results. Solutions of the equation $\Delta_A^{-1}=0$ are called $A$-modes, and solutions of $\Delta_G^{-1}=0$ are referred to as $G$-modes. 

In the extremely prolate system, we can solve the dispersion equations analytically and write the propagators in the simple form
\ba
\Delta_A^{-1}(\omega) &=& k^2(\hat\omega^2-\hat\omega_a^2),
\\
\Delta_G^{-1}(\omega) &=& \frac{\hat\omega^2 k^4}{\hat\omega^2-x^2} (\hat\omega^2-\hat\omega_2^2)(\hat\omega^2-\hat\omega_+^2)(\hat\omega^2-\hat\omega_-^2) ,
\ea
where the dispersion relations are
\ba
\hat\omega_a^2 &=& 1+\frac{m^2}{2k^2},
\\
\hat\omega_2^2 &=&  x^2+\frac{m^2}{2k^2},
\\
2\hat\omega_\pm^2 &=& 1+x^2\pm\sqrt{(1-x^2)(1-x^2+2m^2/k^2)} .
\ea
The modes $\pm\omega_a$, $\pm\omega_2$ and $\pm\omega_+$ are pure real and exist for all wave vectors. The $\pm\omega_-$ modes are either pure real or pure imaginary. When $k$ is greater than a threshold value, which we call $k_{\rm pG}$ (see Eq.~(\ref{k-crit-oblate})), they are real, and when $k$ is less than this threshold they are imaginary and can be written as $\pm i \gamma_a$ with $\gamma_a \in \mathbb{R}$. 

For the oblate distribution, solutions of the dispersion equations can only be found numerically. There are 6 real modes (3 pairs) that exist for all wave vectors which we call $\pm\omega_a$ ($A$-modes), $\pm\omega_-$ and $\pm\omega_+$ ($G$-modes). There are two thresholds (defined in Eq.~(\ref{k-crit-oblate})) below which imaginary modes appear. For $k<k_{\rm oA}$ there is a pair of imaginary $A$-modes $\pm i\gamma_a$, and for $k<k_{\rm oG}$ there is a pair of imaginary $G$-modes $\pm i\gamma_-$. Note that we use the same terminology for prolate and oblate modes without introducing subscripts to distinguish them, but this will not cause confusion because the prolate and oblate systems are considered separately in sections \ref{sec-ex-prolate} and \ref{sec-ex-oblate}. 

\begin{figure}[t]
\hspace{-2cm}
\begin{minipage}{8cm}
\center
\includegraphics[width=1.15\textwidth]{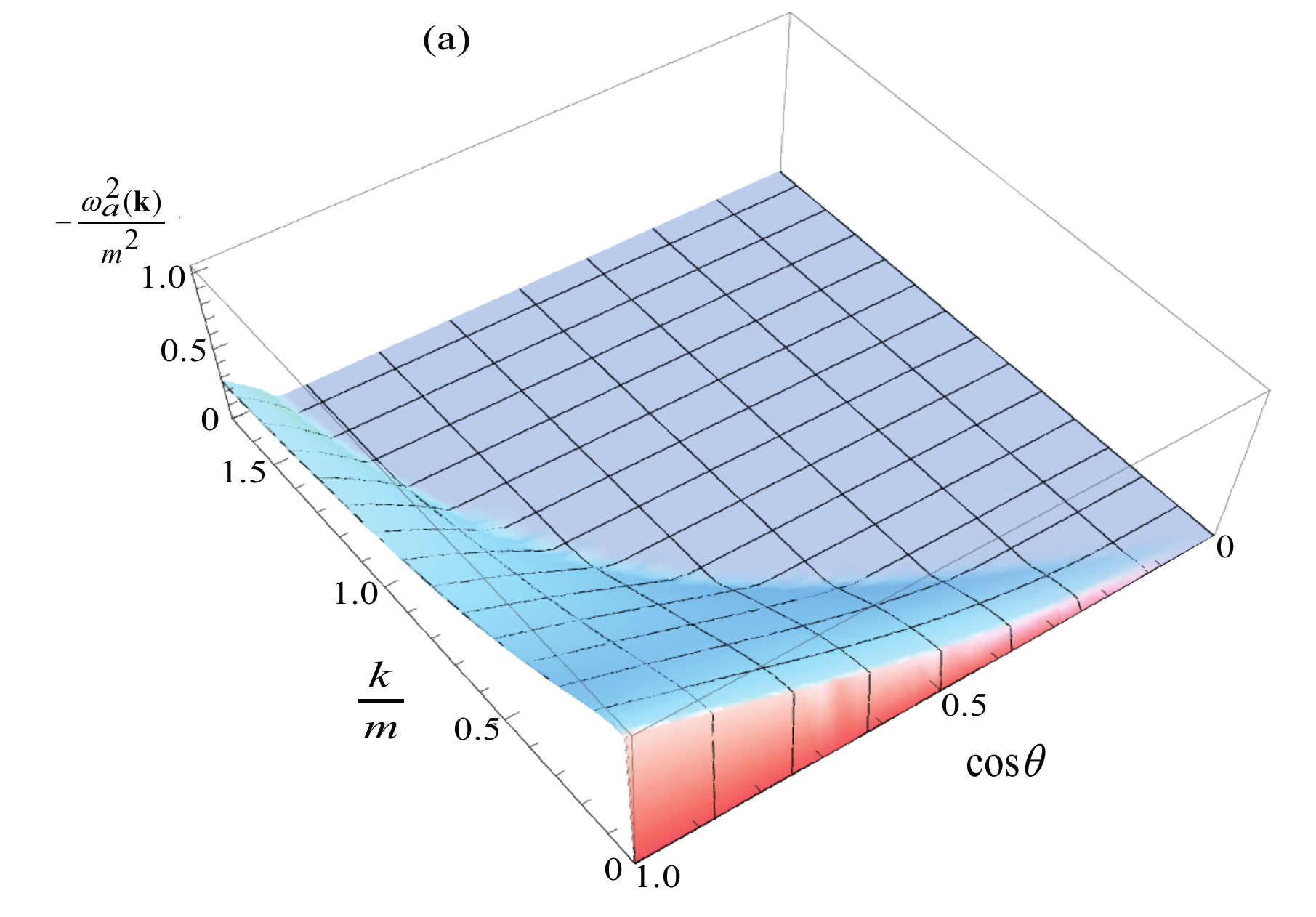}
\end{minipage}
\hspace{10mm}
\begin{minipage}{8cm}
\center
\includegraphics[width=1.15\textwidth]{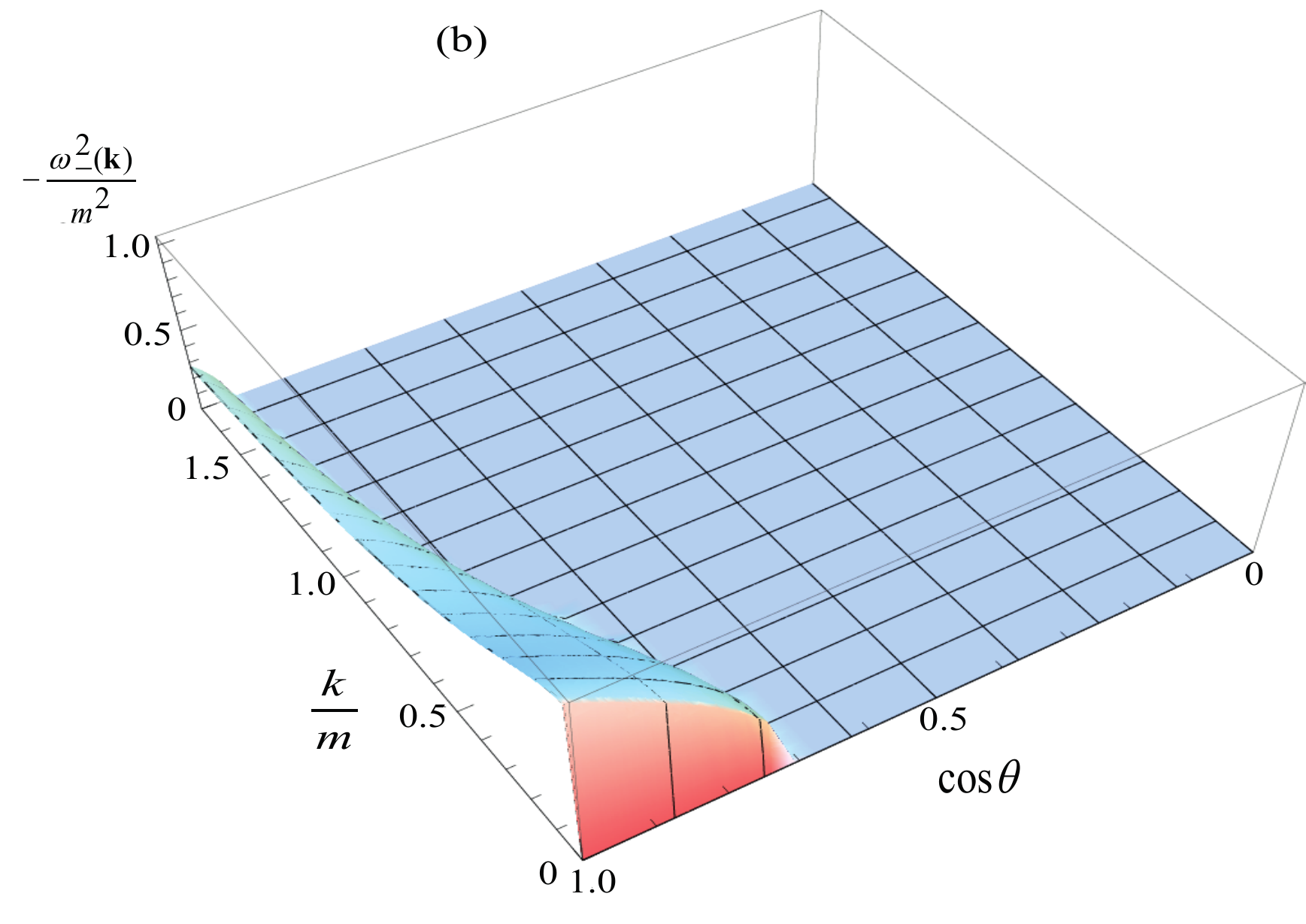}
\end{minipage}
\vspace{-3mm}
\caption{(Color online) Unstable modes for the extremely oblate plasma: $- \omega_a^2$ (panel (a)) and $-\omega_-^2$ (panel (b)) as functions of $k$ and $\cos\theta$ in the domain where the modes exit. The angle $\theta$ is between the vectors ${\bf k}$ and ${\bf n}$.
\label{fig-unstable-modes-oblate}}
\end{figure}

The threshold wave vectors for prolate and oblate systems are \cite{Carrington:2014bla}
\ba
\label{k-crit-prolate}
&& k_{\rm pG} \equiv \frac{m}{\sqrt{2}} |\tan\theta| ,
\\
\label{k-crit-oblate}
~~~~~~~~~~~~~~~~~~~~~~~~~~
&& k_{\rm oA}  \equiv \frac{m}{\sqrt{2}} |\cot\theta| ,
\label{k-crit-oblate}
~~~~~~~~~~~~~
k_{\rm oG}  \equiv \frac{m}{2}\Re \sqrt{\frac{|\cos\theta|\sqrt{\cos^2\theta + 4}+\cos^2\theta-2}{\sin^2\theta}} .
\ea
When $\theta\to \pi/2$ we have $k_{\rm pG}\to\infty$ and the unstable prolate $G$-mode exists for all $k$'s. This behavior is shown in Fig.~\ref{fig-omega-minus-ex-prolate} for the prolate unstable $G$-mode. In the oblate system the situation is reversed and the threshold wave vectors $k_{\rm oA}$ and $k_{\rm oG}$ approach infinity when $\theta\to0$. The unstable $A$ and $G$-modes for the extremely oblate system are shown in Fig.~\ref{fig-unstable-modes-oblate}.

As discussed in Sec.~\ref{sec-form-general}, we calculate the frequency integral of the energy loss formula (\ref{e-loss-Integrand}) with a contour in the lower half plane that encloses all singularities of the integral. The significance of the imaginary modes can be seen immediately. Denoting the real part of an imaginary mode generically as $\gamma$, it is clear that the residue of a pure imaginary mode contains a factor $e^{\gamma t}$ which grows exponentially with time. However,  the magnitude of the unstable mode is small (in mass units), and the region of phase space for which the unstable mode exists is quite limited. Since there is also an oscillatory factor $e^{-i\bar\omega t}$ under the integral  (\ref{e-loss-main}), it is not clear whether the energy loss will increase exponentially as a function of time. 

The integral over the wave vector ${\bf k}$ is taken numerically in spherical coordinates with the $z$-axis along the anisotropy vector ${\bf n}$. Since the integral is ultraviolet divergent, we regulate it by introducing an upper cut-off at some finite momentum $k_{\rm max}$. For both oblate and prolate plasmas, there is a potential divergence when an imaginary mode goes to zero, as the wave vector approaches its threshold value. However,  these divergences cancel exactly (sometimes in combination with the residue from the pole at $\omega=0$). There are divergences that depend on the azimuthal angle when $\bar\omega= 0$ and $\bar\omega= \pm\omega_-$ but they are odd and can be regulated using a principal part prescription. 

The current contribution to the energy loss, or the energy loss with the uncorrelated initial condition, is very oscillatory and hard to calculate, but we have checked that it is of the order of the equilibrium energy loss discussed in Sec.~\ref{sec-stable} and its magnitude is much smaller than the field contribution. One example of this current contribution is given in Sec.~\ref{sec-ex-prolate}. 

In the two subsequent sections we present our numerical results on the energy loss in the extremely prolate and extremely oblate plasmas. In all our numerical calculations $C_R=3$, which corresponds to a gluon, and our results are expressed in the units of $m$. As in section \ref{sec-stable}, the energy loss is divided by $g^2 m^2$ and therefore the value of the coupling constant $g$ is not specified.

\subsection{Extremely prolate plasma}
\label{sec-ex-prolate}

\begin{figure}[t]
\hspace{-7mm}
\begin{minipage}{8cm}
\center
\includegraphics[width=1.0\textwidth]{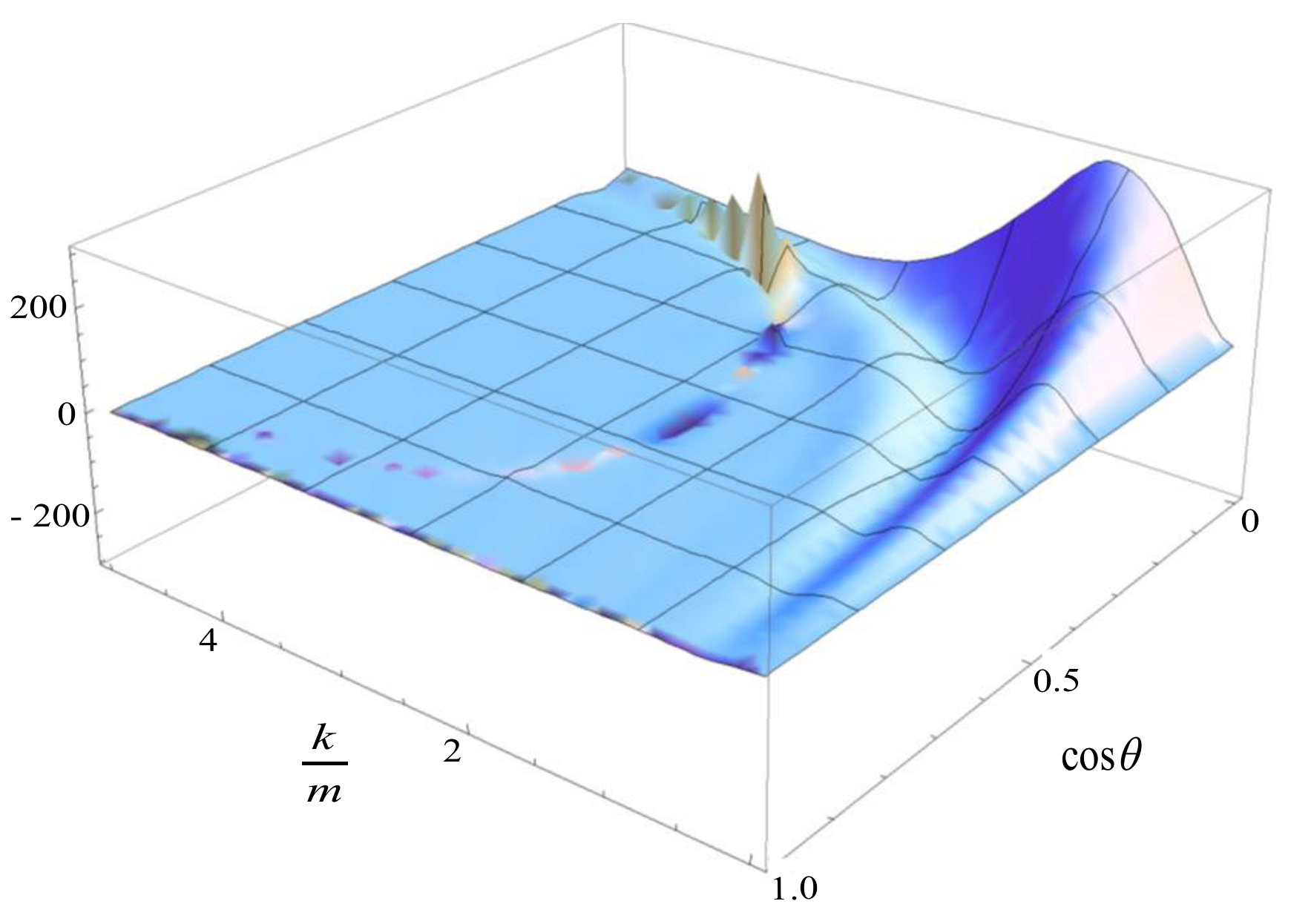}
\vspace{-2mm}
\caption{(Color online) The integrand of the energy loss with $\cos\alpha=1$ in extremely prolate plasma as a function of $k$ and $\cos\theta$ for $\Theta=\pi/12$ and $t=8/m$. The integral over azimuthal angle $\phi$ has been performed.}
\label{fig-integrand-prolate}
\end{minipage}
\hspace{3mm}
\begin{minipage}{9cm}
\center
\includegraphics[width=1.03\textwidth]{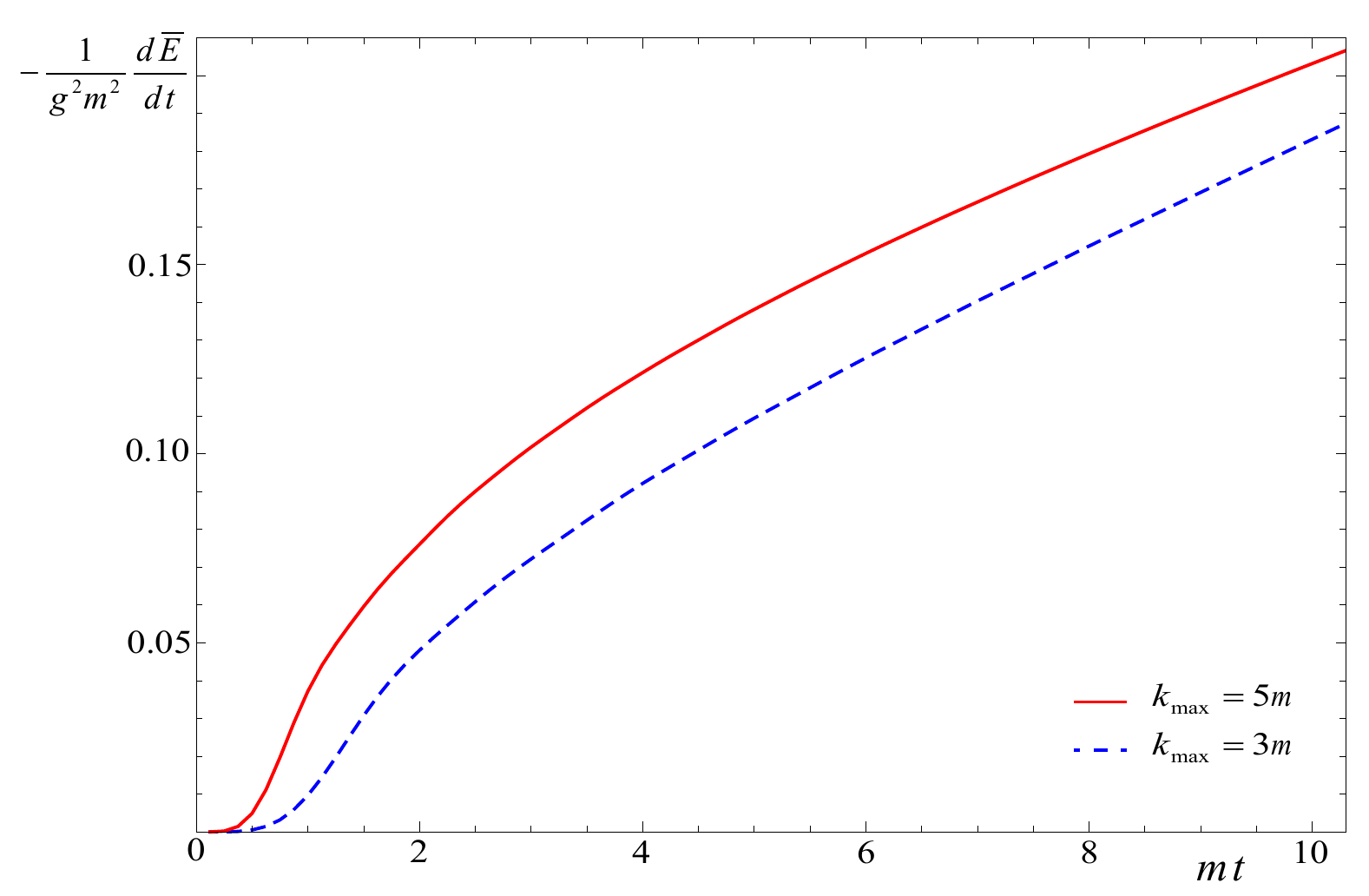}
\vspace{-8mm}
\caption{(Color online) Current contribution to the energy loss as a function of $t$ for $\Theta=0$, for two different choices of the $k_{\rm max}$. The red (solid) curve is for $k_{\rm max}= 5m$, and the blue (dashed) curve is $k_{\rm max}=3m$.}
\label{fig-el-prolate-current}
\end{minipage}
\end{figure}

In Fig.~\ref{fig-integrand-prolate} we present the integrand of the energy loss in prolate plasma as a function of $k$ and $\cos\theta$ for $\Theta=\pi/12$. The integral over azimuthal angle $\phi$ has been done, and the small spikes at the top of the figure are produced by numerical issues.  The meaning of the angles $\theta, \; \phi,\; \Theta$ is explained by Eq.~(\ref{coordinate-system}). Comparing this plot to that shown in Fig.~\ref{fig-omega-minus-ex-prolate}, one clearly sees the influence of the unstable mode - the integrand is large in the domain of $k$ and $\cos\theta$ where the mode $\omega_-$ exists. 

Fig.~\ref{fig-el-prolate-current} shows the current contribution to the energy loss (or the energy loss with $\cos\alpha=0$) as a function of time for $\Theta=0$. Since $d\overline{E}/dt$ is negative, the parton loses energy. The two curves represent two values of $k_{\rm max} = 3m$ and $k_{\rm max} = 5m$, and one sees that the magnitude of the energy loss increases with $k_{\rm max}$. The result is time dependent, but it is approximately of the same magnitude as the equilibrium energy loss at a given $k_{\rm max}$. As seen in Fig.~\ref{fig-eq}, the equilibrium energy loss equals $-0.12 \, g^2m^2$ for   $k_{\rm max}= 3m$ and $-0.18 \, g^2m^2$ for $k_{\rm max}= 5m$.

\begin{figure}[t]
\hspace{-4mm}
\begin{minipage}{8.5cm}
\center
\includegraphics[width=1.0\textwidth]{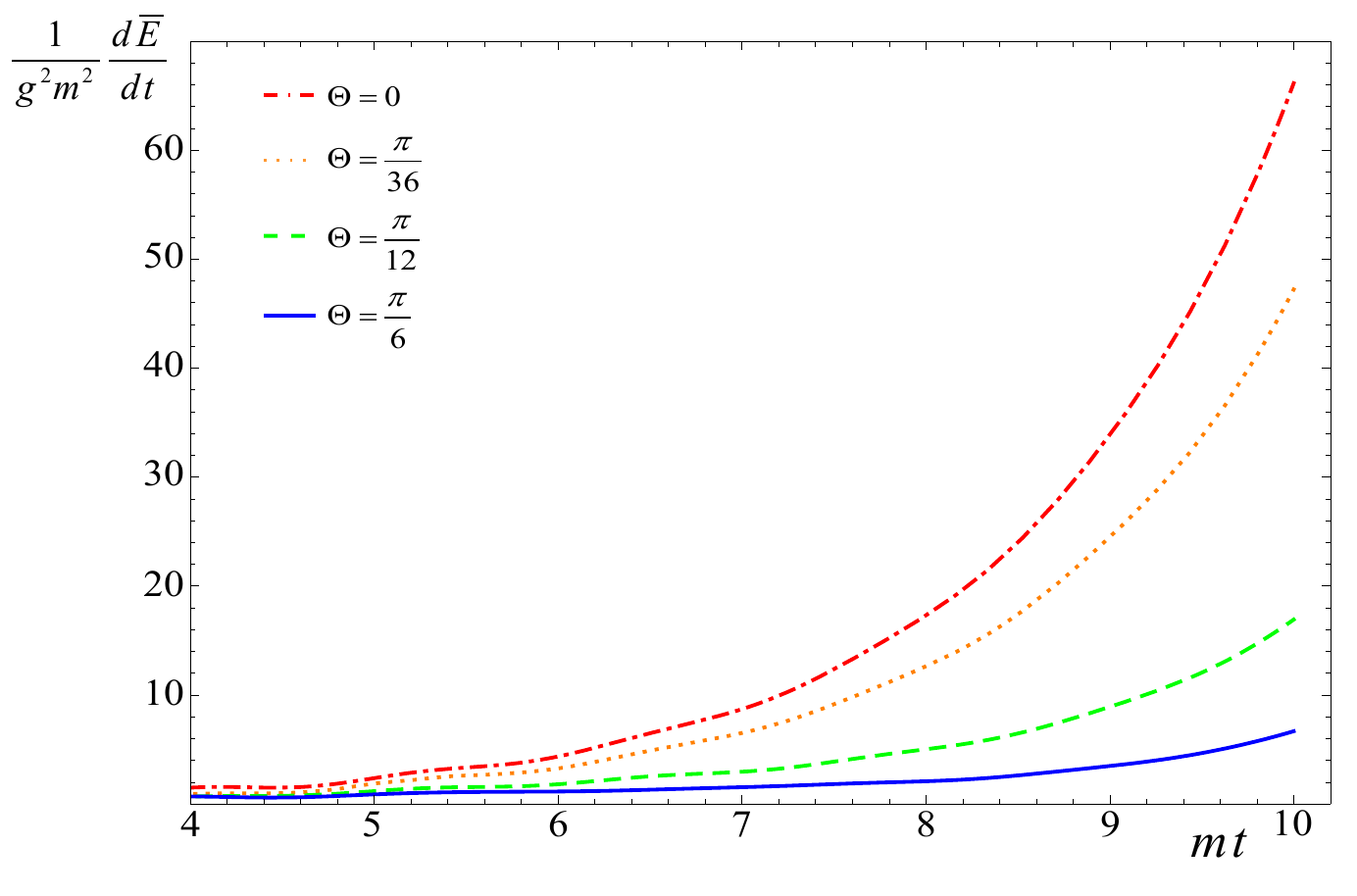}
\vspace{-7mm}
\caption{(Color online) The field contribution to the energy loss in prolate plasma as a function of time for four angles $\Theta$ between the parton velocity and the anisotropy direction: $\Theta=0$ (red, dot-dashed), $\Theta=\pi/36$ (orange, dotted), $\Theta=\pi/12$ (green, dashed), and $\Theta=\pi/6$ (blue, solid). The cut-off parameter is $k_{\rm max}= 5 m$. 
\label{fig-prolate-angles}}
\end{minipage}
\hspace{1mm}
\begin{minipage}{9.2cm}
\center
\vspace{-5mm}
\includegraphics[width=1.0\textwidth]{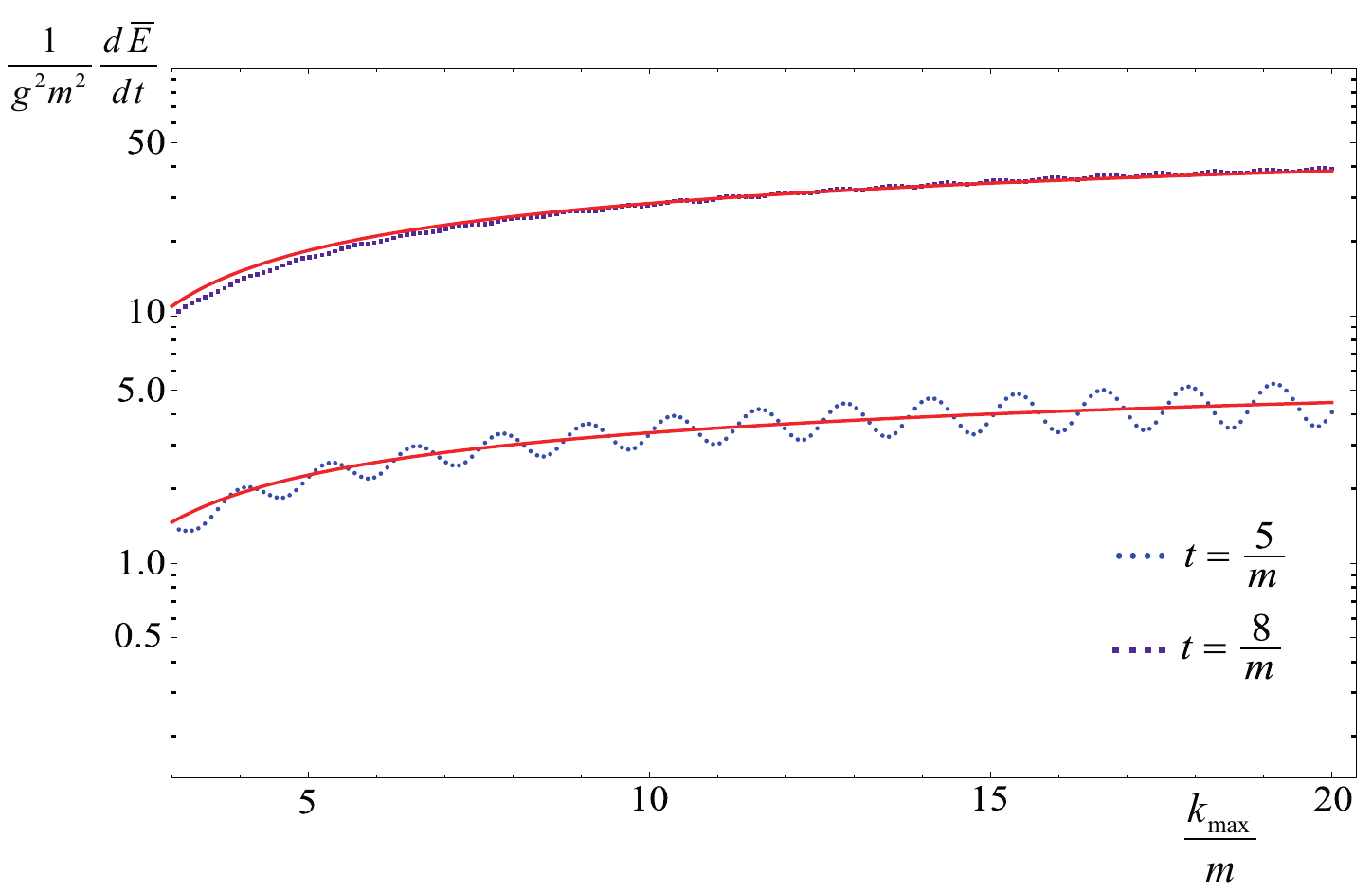}
\vspace{-6mm}
\caption{(Color online)  The field contribution to the energy loss in prolate plasma at $\Theta=0$ as a function of $k_{\rm max}$ for $t=5/m$ (circles) and $t=8/m$ (squares). The red (solid) curves are a logarithmic fit.}
\label{fig-prolate-kmax}
\end{minipage}
\end{figure}

In Fig.~\ref{fig-prolate-angles} we show the field contribution to the energy loss (with $\cos\alpha=1$) as a function of time for $k_{\rm max}= 5 m$ and four angles $\Theta$ between the parton velocity and the anisotropy vector ${\bf n}$. The energy loss  $d\overline{E}/dt$ is positive and it increases exponentially with time, showing the effect of the unstable modes. The parton thus gains the energy and the magnitude of $d\overline{E}/dt$ at later times is much bigger than in equilibrium plasmas (see Fig. \ref{fig-eq}). The sign of the field contribution to the energy loss is determined by the sign of the phase factor $\cos\alpha$, and therefore if we change the initial condition from $\cos\alpha=1$ to $\cos\alpha=-1$ we will get exponentially growing energy loss instead of exponentially growing energy gain. Since the field contribution to the energy loss is much bigger than the current contribution, the sign of $d\overline{E}/dt$ is actually controlled by the sign of $\cos\alpha$. Therefore, the energy loss crucially depends on the initial condition. 

\begin{figure}[b]
\centering
\includegraphics[width=0.60\textwidth]{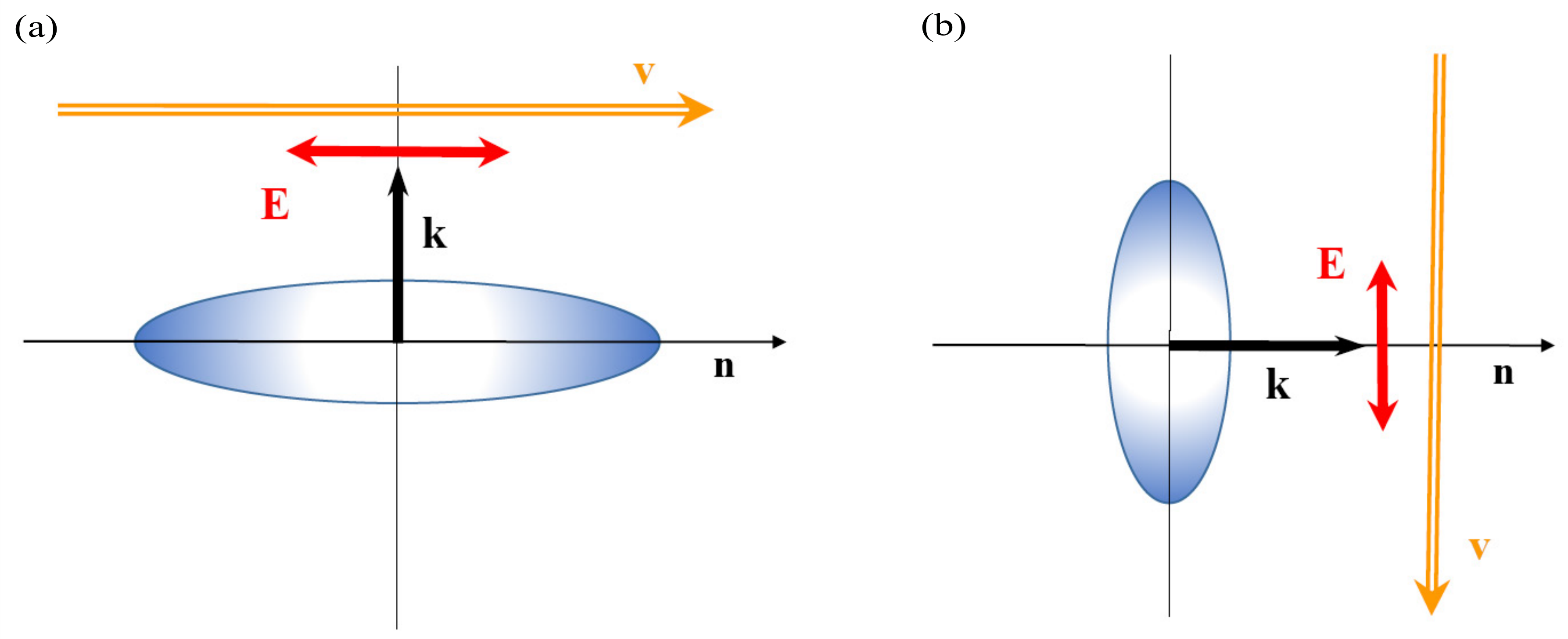}
\caption{(Color online) The configurations when the energy loss is maximal in the prolate (a) and oblate (b) plasmas. The arrows labeled with ${\bf n},\; {\bf k}, \;{\bf E}$ and ${\bf v}$ denote, respectively, the anisotropy vector, the wave vector of the most important unstable mode, the electric field associated with this mode, and the velocity of the test parton.}
\label{fig-configurations}
\end{figure}

One observes in Fig.~\ref{fig-prolate-angles} a strong directional dependence of the energy loss.  For a prolate system, the most important wave vectors are those for which ${\bf k}\perp{\bf n}$, where the threshold wave vector (\ref{k-crit-prolate}) goes to infinity, see also Fig.~\ref{fig-omega-minus-ex-prolate}. When ${\bf k}\perp{\bf n}$ the unstable mode has an associated electric field that is parallel to the vector ${\bf n}$. This point is explained in Appendix \ref{sec-x0and1}. The energy transfer is most efficient when the electric field is parallel to the  velocity of the test parton (${\bf v}\parallel {\bf E}$). Therefore, one expects the largest energy transfer when ${\bf v}\parallel {\bf n}$. This argument is shown schematically in Fig. \ref{fig-configurations}a and verified by the results presented in Fig.~\ref{fig-prolate-angles} which demonstrates that the magnitude of the energy loss is maximal at $\Theta=0$, and rapidly decays when the angle $\Theta$ grows. 

In Fig.~\ref{fig-prolate-kmax} we show (in a logarithmic scale) the energy loss as a function of $k_{\rm max}$ for $\Theta=0$ and two times: $t=5/m$ and $t=8/m$. The energy loss oscillates slightly, but the $k_{\rm max}$ dependence can be roughly approximated as $\log k_{\rm max}$, as in the equilibrium case. As discussed in Sec.~\ref{sec-stable}, the divergence at large $k_{\rm max}$ indicates a breakdown of the classical theory.

\begin{figure}[t]
\centering
\includegraphics[width=0.55\textwidth]{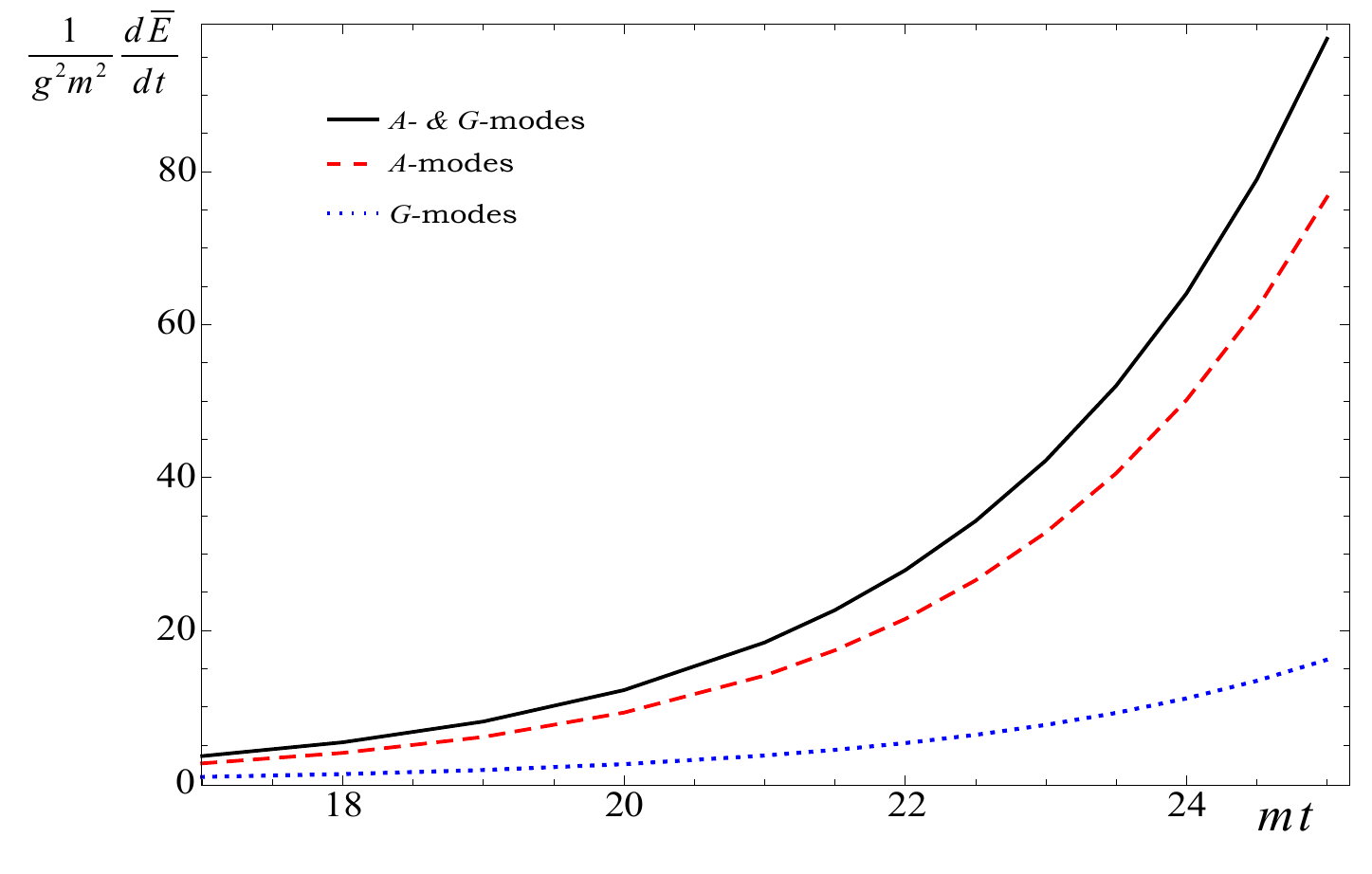}
\vspace{-5mm}
\caption{(Color online) The field contribution to the energy loss in oblate plasma as a function of time for $\Theta = \pi/2$. The red (dashed) line corresponds to the effect of $A$-modes, blue (dotted) line is the $G$-modes, and the black (solid) line represents the sum. }
\label{fig-el-oblate-1}
\end{figure}

\subsection{Extremely oblate plasma}
\label{sec-ex-oblate}

Calculations in oblate plasma are much more difficult than those in prolate plasma because the components of the polarization tensor defined by Eq.~(\ref{pi-decomposition}) have a more complicated structure. They contain square roots that are not defined along the section of the real axis where the arguments of the roots are negative. There are therefore contributions to the frequency integral from the discontinuities between the upper and lower sides of the cuts that are difficult to calculate. We have checked for several cases that they are small when compared to the pole contributions and we therefore neglect them. One consequence of this more complicated structure is that the spectrum of collective modes is richer - there two unstable modes instead of one as in the case of prolate plasma. It is impossible to solve the dispersion equations analytically, and one can only obtain the dispersion relations numerically. Finally, there is a technical complication related to the fact that the dominant contribution to the energy loss in the oblate plasma comes from the domain of wave vectors ${\bf k}$ which are almost parallel to the anisotropy vector ${\bf n}$. When ${\bf k}\parallel {\bf n}$ we have ${\bf n}_T=0$ and the decomposition (\ref{Sigma-A-B-C-D}) is ill defined. This occurs because, when ${\bf k} ||{\bf n}$, the matrix $\Sigma$ does not depend on two independent vectors ${\bf k}$ and ${\bf n}$ but only on one vector ${\bf k}$ or ${\bf n}$. Consequently, the decomposition (\ref{Sigma-A-B-C-D}) should include only two terms with the matrices $A$ and $B$. The propagator has the form 
\ba
\label{oblatex1a}
\Delta^{ij} (\omega) &=& \frac{1}{\omega^2-{\bf k}^2-\alpha(\omega)} \, A^{ij}
+ \frac{1}{\omega^2 -\beta(\omega)} B^{ij} 
\ea
and the components of the polarization tensor are easily calculated in this special case as (see \cite{Carrington:2014bla} for details):
\ba
\label{oblatex1b}
\alpha(\omega) = \frac{m^2}{2}-\frac{m^2(\omega^2-k^2)}{4\omega^4},~~~~~~~
\beta(\omega) = \frac{m^2}{2} \,.
\ea
In our calculation the domain where ${\bf k} ||{\bf n}$ was treated analytically and combined with the results of the numerical computation, as described below. Because of these technical difficulties, we give numerical results for the extremely oblate plasma only for a rather small value of $k_{\rm max}=2m$. The equilibrium energy loss for this value of $k_{\rm max}$, which will be used as a reference point, equals $-0.079\, g^2m^2$, see Fig.~\ref{fig-eq}.

As in the case of prolate plasma, the current contribution is significantly smaller than the field contribution. The latter is shown in Fig.~\ref{fig-el-oblate-1}  when the parton's momentum is perpendicular to the anisotropy vector ${\bf n}$ ($\Theta = \pi/2$). The red (dashed) line represents the contribution due to the $A$-modes, blue (dotted) line represents the $G$-modes and black (solid) one gives the sum. The black (solid) is not exactly the sum of the red (dashed) and blue (dotted) because in the calculation with all modes the points at $x=\pm 1$, which are obtained analytically, are combined and integrated together with the numerical data which is calculated over the range $-0.9996<x<0.9996$. One observes in Fig.~\ref{fig-el-oblate-1} that the unstable $A$-mode is responsible for the largest effect. Since the field contribution to the energy loss is much bigger than the current contribution, the sign of the energy loss is determined by the sign of $\cos\alpha$ which expresses the dependence on the initial conditions. $d\overline{E}/dt$ is negative for $\cos\alpha <0 $ and it is positive when $\cos\alpha > 0 $. As seen in Fig.~\ref{fig-el-oblate-1}, the energy loss in oblate plasma can be orders of magnitude bigger than in an equilibrium plasma with the same $k_{\rm max}$.

\begin{figure}[t]
\hspace{-7mm}
\begin{minipage}{8.5cm}
\center
\includegraphics[width=1.07\textwidth]{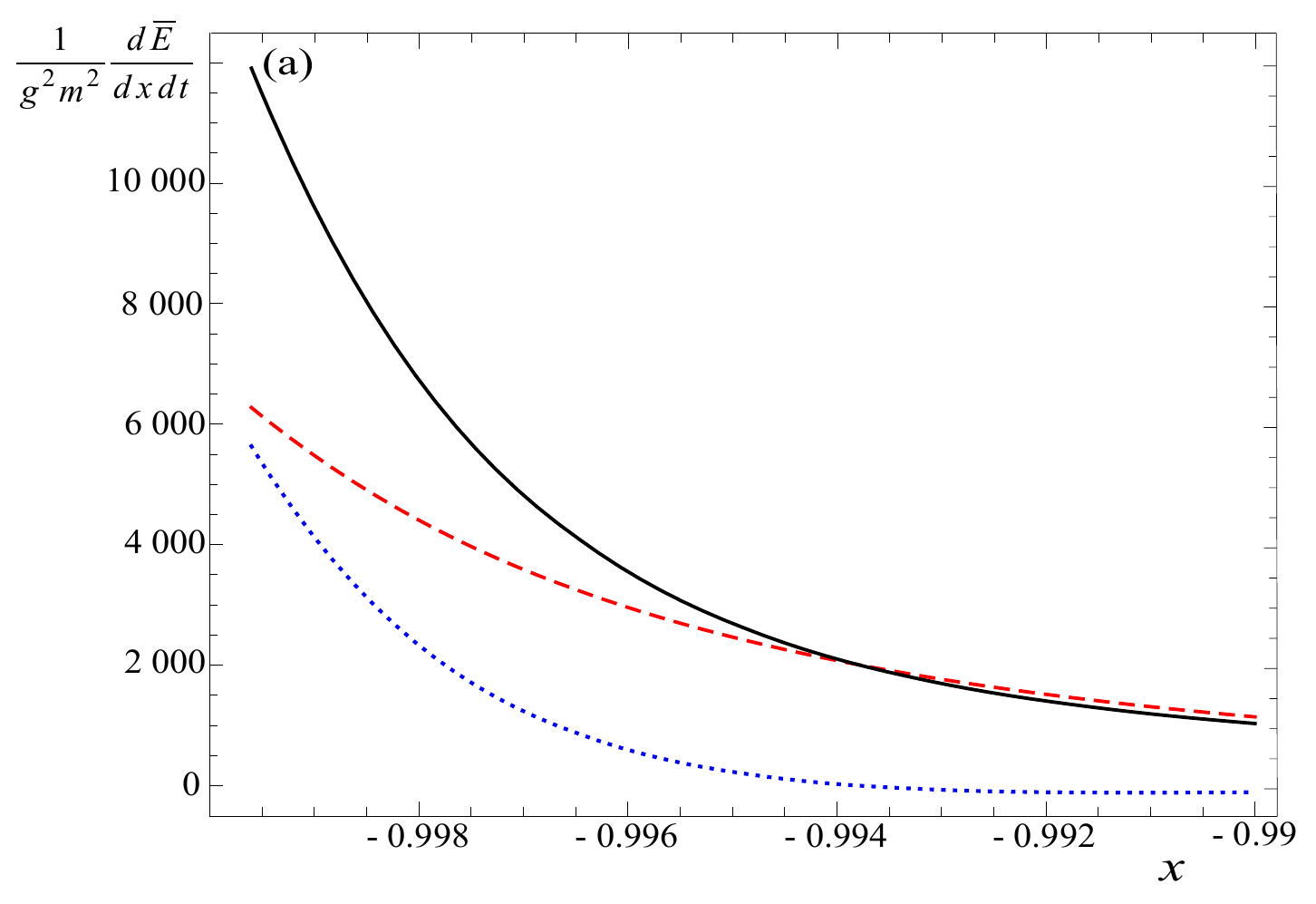}
\end{minipage}
\hspace{2mm}
\begin{minipage}{8.5cm}
\center
\includegraphics[width=1.09\textwidth]{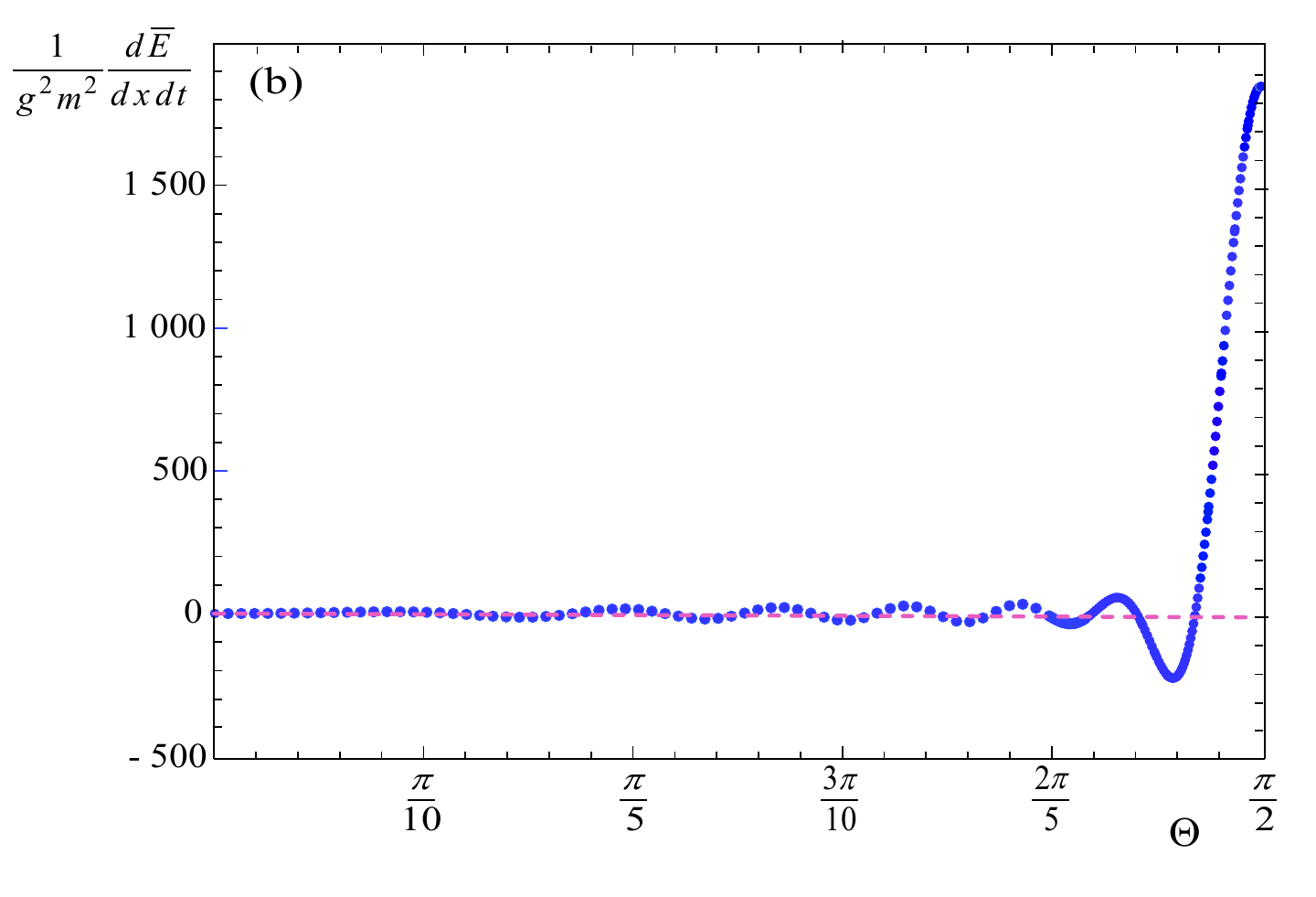}
\end{minipage}
\vspace{-4mm}
\caption{(Color online) The integrand of the field contribution to oblate energy loss after integrating over $\phi$ and $k$ for $t=25/m$. In the panel (a) the integrand is shown as a function of $x \equiv \cos\theta$ for $\Theta=\pi/2$ and in the panel (b) as a function of $\Theta$ for $x=1$. The red (dashed), blue (dotted) and black (solid) lines in the panel (a) represent the effect of, respectively, $A$-modes, $G$-modes and the sum of $A$ plus $G$-modes.}
\label{fig-el-oblate-2}
\end{figure}

\begin{figure}[b]
\centering
\includegraphics[width=0.55\textwidth]{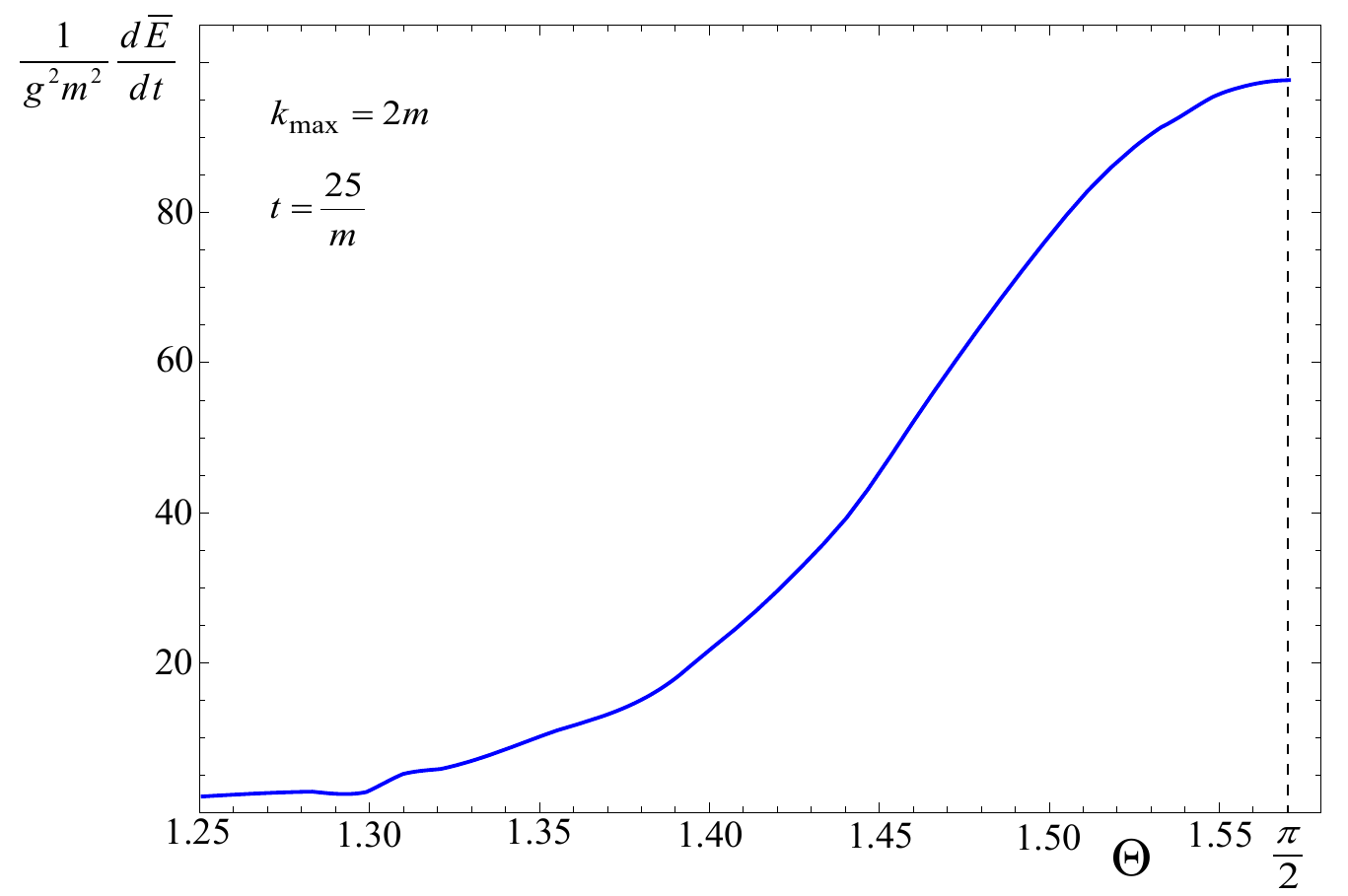}
\vspace{-2mm}
\caption{(Color online) The field contribution to the energy loss in oblate plasma as a function of the angle $\Theta$ for $t=25/m$. }
\label{fig-el-oblate-3}
\end{figure}

For an extremely oblate system, the most important wave vectors are those for which ${\bf k}\parallel{\bf n}$, since both of the thresholds $k_{\rm oA}$ and $k_{\rm oG}$ go to infinity in this limit, see Eq.~(\ref{k-crit-oblate}). This behavior is also shown in Fig \ref{fig-unstable-modes-oblate}. As explained in Appendix~\ref{sec-x0and1}, instead of two different pairs of imaginary modes  $A$ and $G$, we have two pairs of identical modes which are purely transverse when ${\bf k}\parallel{\bf n}$. The electric field associated with these modes is perpendicular to both ${\bf k}$ and ${\bf n}$. Since the energy loss is maximal when the parton velocity is parallel to the electric field,  such a situation occurs in the oblate system when ${\bf v}$ is perpendicular to ${\bf n}$, or $\Theta=\pi/2$. This argument is shown schematically in Fig. \ref{fig-configurations}b. The effect is seen explicitly in Figs.~\ref{fig-el-oblate-2} and \ref{fig-el-oblate-3}. The left panel of Fig. \ref{fig-el-oblate-2} shows that for both $A$- and $G$-modes the energy loss is dominated by the region $x\approx 1$, and the right panel proves that when $x=1$ the biggest effect is observed when $\Theta=\pi/2$. Fig. \ref{fig-el-oblate-3} presents the energy loss as a function of $\Theta$ for $t=25/m$. The figure shows that $d\overline{E}/dt$ drops rapidly when $\Theta$ becomes smaller than $\pi/2$. 


\section{Discussion and outlook}
\label{sec-conclusions}

Let us first summarize our study. We have derived the energy loss formula for a high-energy parton flying across an unstable plasma which experiences a rapid temporal evolution due to exponentially growing collective modes. Since the formula is found as the solution of an initial value problem, initial values of the chromodynamic fields present in the plasma must be chosen. Except in special cases, the energy loss formula includes an effect of self-interaction which must be subtracted to get a physically meaningful result.  In the case of equilibrium plasmas, the initial conditions are `forgotten,' and the well-known formula of collisional energy loss is reproduced.  When the initial conditions are chosen in such a way that the initial fields are not correlated with the current generated by the test parton, the parton typically looses energy, and the magnitude of the energy loss is comparable to that in an equilibrium plasma of the same mass $m$ (\ref{Debye-mass}). When the initial chromodynamic field is induced by the parton, it can be either accelerated or decelerated depending on the relative phase factor. With correlated initial conditions, the magnitude of the energy loss grows exponentially in time and can much exceed the absolute value of the energy loss in an equilibrium plasma. 

We have derived an expression for the energy loss for arbitrarily prolate or oblate plasmas, and performed numerical calculations for the specific examples of the  extremely prolate and extremely oblate systems. The energy loss is not only time dependent but it is also strongly directionally dependent. The configurations when the energy loss is maximal in the prolate and oblate plasmas are illustrated in Fig.~\ref{fig-configurations}. In these special configurations, the magnitude of the energy loss can be much bigger than that in an equilibrium plasma. Beyond a narrow cone which is centered around the optimal direction, the energy loss rapidly drops. 

It is interesting to consider the possible consequences of our findings for the jet suppression observed in relativistic heavy-ion collisions. Since a high-energy parton can be either accelerated or decelerated in an unstable plasma, we expect that the energy loss strongly fluctuates, and that the fluctuations are particularly large in the configurations depicted in Fig.~\ref{fig-configurations}. Quark-gluon plasma at an early stage of a relativistic heavy-ion collision has initially a prolate momentum distribution which evolves fast due to free streaming to an oblate momentum distribution. During the process of equilibration the plasma is oblate and it remains oblate in the  subsequent evolution because of viscosity effects \cite{Florkowski:2013lya}.  Jet quenching is observed at both RHIC and LHC at almost vanishing rapidity in the center of mass of colliding nuclei. This configuration is just as shown in Fig.~\ref{fig-configurations}b, where the jet momentum is transverse to the vector ${\bf n}$. We suspect that the jet quenching pattern can be changed when the jet axis is tilted in such a way that the near-side jet has a small but positive (negative) rapidity while the away-side jet has a small but negative (positive) rapidity. The effect of unstable modes is then reduced and the energy-loss fluctuations are expected be much smaller. 

One should remember that we have discussed here only collisional energy loss. There are simple arguments that indicate that  radiative energy loss behaves very differently \cite{Majumder:2009cf}. Radiative energy loss is controlled by the parameter $\hat{q}$ which measures the momentum broadening of a parton. This parameter is by definition positive and grows exponentially in an unstable plasma, as does the radiative energy loss, which is always negative. Therefore, before we draw a conclusion about the possible role of unstable plasma in jet suppression phenomenology, the effects of both collisional and radiative energy loss must be combined. This requires a computation of $\hat{q}$ in unstable plasmas which we plan to publish soon. 

\section*{Acknowledgments}

This work was partially supported by the Polish National Science Centre under grant 2011/03/B/ST2/00110 and the Natural Sciences and Engineering Research Council of Canada.

\appendix

\section{Reality of energy loss}
\label{sec-reality-EL}

We prove here that the energy loss (\ref{e-loss-main}) is real for any momentum distribution that satisfies the mirror symmetry $f({\bf p})=f(-{\bf p})$. For this purpose we take the complex conjugate of the formula (\ref{e-loss-main}) and obtain
\ba
\label{e-loss-main-complex-1}
\frac{dE^*(t)}{dt} = - g Q^a v^i 
\int {d^3k \over (2\pi)^3}
\int_{-\infty - i\sigma}^{\infty - i\sigma}
{d\omega \over 2\pi i}
e^{ i(\omega -\bar{\omega})t}
\Delta^{* ij}(\omega,{\bf k})
\Big[ 
- \frac{ i \omega g Q^a v^j}{\omega - \bar{\omega}}
+ \epsilon^{jkl} k^k B_{0a}^{*l} ({\bf k})
- \omega D_{0a}^{*j} ({\bf k}) \Big], 
\ea
where the inverse matrix $\Sigma^{-1}$ is replaced by the retarded propagator $\Delta$. Now we change the integration variables $\omega \rightarrow - \omega$ and ${\bf k} \rightarrow - {\bf k}$ which gives
\ba
\label{e-loss-main-complex-1}
\frac{dE^*(t)}{dt} = - g Q^a v^i 
\int {d^3k \over (2\pi)^3}
\int_{- \infty + i\sigma}^{\infty + i\sigma}
{d\omega \over 2\pi i}
e^{- i(\omega -\bar{\omega})t}
\Delta^{* ij}(-\omega,-{\bf k})
\Big[ -
\frac{i\omega g Q^a v^j}{\omega - \bar{\omega}}
- \epsilon^{jkl} k^k B_{0a}^{*l} (-{\bf k})
+ \omega D_{0a}^{*j} (-{\bf k}) \Big]\,.
\ea
Since the initial fields ${\bf B}_{0a}({\bf r})$ and ${\bf D}_{0a}({\bf r})$ are pure real in coordinate space, we have 
\ba
\label{sym2}
{\bf B}_{0a}({\bf k}) = {\bf B}^*_{0a}(-{\bf k}),~~~~~~{\bf D}_{0a}({\bf k}) = {\bf D}^*_{0a}(-{\bf k}).
\ea
In our study \cite{Carrington:2014bla} we have proven that for mirror-symmetric momentum distributions, the retarded propagator defined by Eq.~(\ref{prop}) satisfies the relations
\ban
\Re\Delta^{ij}(-\omega,-{\bf k}) = \Re\Delta^{ij}(\omega,{\bf k}) , ~~~~~~~~~~
\Im\Delta^{ij}(-\omega,-{\bf k}) = -\Im\Delta^{ij}(\omega,{\bf k}) ,
\ean
which give
\ba
\label{sym1}
\Delta^{*ij}(-\omega,-{\bf k}) =\Delta^{ij}(\omega,{\bf k}) .
\ea
Using the relations (\ref{sym2}) and (\ref{sym1}), the right side of Eq.~(\ref{e-loss-main-complex-1}) is identical to the right side of Eq.~(\ref{e-loss-main}), which completes the proof that the energy loss given by the formula (\ref{e-loss-main}) is real. 

\section{Temporal axial and Feynman-Lorentz gauges}
\label{sec-gauge}

In this appendix we show that the temporal axial gauge is particularly convenient for the energy loss calculation because it naturally provides a gauge independent expression for the energy loss that depends only on the electric and magnetic fields. In contrast, current conservation must be explicitly enforced in Feynman-Lorentz gauge. To simplify the problem, we consider here an electromagnetic plasma.

The electromagnetic analog of the energy loss formula (\ref{e-loss-main}) is clearly gauge invariant, as the derivation of the formula is gauge invariant at every step. The gauge dependent potential $A^\mu$ is not used at all, and the energy loss is written in a form that depends only on the dielectric tensor and electric and magnetic fields, which are physical quantities. However, when we switch to the terminology of quantum field theory and the inverse dielectric tensor is replaced by the photon propagator in the temporal axial gauge, the gauge independence of the formula (\ref{e-loss-main}) is not evident any more. In this section we will show that although the energy loss formula looks different in the Feynman-Lorentz gauge, it is still gauge invariant. We also explain why temporal axial gauge is much more convenient for the energy loss calculation. In this appendix we use the usual (two-sided) Fourier transformation and not the one-sided transformation, which was used in Sec.~\ref{sec-form-general}.

To further simplify the problem we will consider not the whole energy loss formula but only the electric field generated by the test particle in vacuum. We will solve the Maxwell equation
\be
\label{EoM-Maxwell}
\partial_\mu F^{\mu \nu}(x) = j^\nu(x),
\ee
where $x = (t,{\bf r})$, $F^{\mu \nu} \equiv \partial^\mu A^\nu - \partial^\nu A^\mu$ and $j^\mu$ is the particle's current. The electric field, which is the physical quantity of interest, is expressed through the four-potential as
\be
\label{E-field-def}
{\bf E}(x) =  - \nabla A^0(x) -  \dot{\bf A}(x) ,
\ee
which in momentum space is
\ba
\label{E-field-def2}
{\bf E}(k) = - i {\bf k}  A^0(k) + i \omega {\bf A}(k) .
\ea
We note that $k=(\omega, {\bf k})$ denotes here the four-vector and not $|{\bf k}|$.

In order to solve equation (\ref{EoM-Maxwell}) for the potential, one must choose a gauge. The resulting solution has the form $A$ = propagator $\times$ current. Both the propagator and the vector potential are gauge dependent. However, if we calculate the electric field from the potential using Eq.~(\ref{E-field-def}) or (\ref{E-field-def2}), the result must be gauge independent. This is true when current conservation is imposed. 

We start by considering Feynman-Lorentz gauge ($\partial_\mu A^\mu =0$) in which the Maxwell equation (\ref{EoM-Maxwell}) is 
\be
\label{EoM-Maxwell-2}
 \Box A^\mu(x) = j^\nu(x),
\ee
and the (two-sided) Fourier transformed solution reads
\be
\label{prop-LOR-1}
A^\mu(k) = \Delta^{\mu \nu}_{\rm FLG}(k) j_\nu(k) ,
\ee
where
\be
\label{prop-LOR}
\Delta^{\mu \nu}_{\rm FLG}(k) = - \frac{g^{\mu \nu}}{k^2 + i{\rm sgn(\omega)}0^+} = g^{\mu \nu} D_{\rm FLG}(k)
\ee
is the retarded photon propagator in the Feynman-Lorentz gauge. From equations (\ref{E-field-def2}), (\ref{prop-LOR-1}) and (\ref{prop-LOR}) we obtain the electric field generated by the current $ j^\nu(k)$ 
\be
\label{E-field-LG-1}
E^i(k) = - i D_{\rm FLG}(k) \big( k^i  j^0(k) - \omega j^i(k) \big) .
\ee

Now we consider the temporal axial gauge ($A^0=0$). The (two-sided) Fourier transformed field equation (\ref{EoM-Maxwell}) splits into two equations
\ba
\label{TAG-eq-1}
- \omega  k^i A^i (k) &=& j^0(k) ,
\\[2mm]
\label{TAG-eq-2}
\big[(- \omega^2 + {\bf k}^2)  \delta^{ij} - k^i k^j ] A^j (k) &=&  j^i(k) .
\ea
The solution of the second equation (\ref{TAG-eq-2}) is
\be
\label{sol-TAG}
 A^i (k) = - \Delta^{ij}_{\rm TAG}(k) j^j(k) ,
\ee
where
\be
\label{prop-TAG}
\Delta^{ij}_{\rm TAG}(k) = 
\frac{1}{\omega^2 + i{\rm sgn}(\omega)0^+} \frac{k^ik^j}{{\bf k}^2}
+ \frac{1}{\omega^2 -{\bf k}^2 + i{\rm sgn}(\omega)0^+}
\Big(\delta^{ij} - \frac{k^ik^j}{{\bf k}^2}\Big) 
\ee
is the retarded photon propagator in  the temporal axial gauge.  Substituting the solution (\ref{sol-TAG}) into Eq.~(\ref{E-field-def2}) and using ($A^0=0$) we obtain
\be
\label{E-field-TAG}
 E^i (k) = -  i \omega \Delta^{ij}_{\rm TAG}(k) j^j(k) .
\ee

We have found that in Feynman-Lorentz gauge the electric field is given by Eq.~(\ref{E-field-LG-1}) and in temporal axial gauge it is given by Eq.~(\ref{E-field-TAG}) with the additional constraint (\ref{TAG-eq-1}). The two equations for the electric field look different, but if current conservation is imposed they are in fact the same. Current conservation gives the relation 
\be
\label{curr-cons-k}
\omega j^0(k) = {\bf k} \cdot {\bf j}(k) \,.
\ee 
Using (\ref{curr-cons-k}) the electric field obtained from the Feynman-Lorentz gauge (\ref{E-field-LG-1}) can be written in the form
\be
\label{E-field-LG-2}
E^i(k) = -\frac{i}{\omega} D_{\rm FLG}(k) \big( k^i  k^j  - \omega^2 \delta^{ij}  \big)  j^j (k) \,.
\ee
Equations (\ref{prop-LOR}) and (\ref{prop-TAG}) give the equality 
\be
\label{props-both}
\Delta^{ij}_{\rm TAG}(k) = \frac{1}{\omega^2} D_{\rm FLG}(k) \big( k^i  k^j  - \omega^2 \delta^{ij}  \big) .
\ee
Using Eq.~(\ref{props-both}) it is easy to see that the expressions (\ref{E-field-TAG}) and (\ref{E-field-LG-2}) are equivalent. 

When working in the temporal axial gauge, current conservation merely tells us that the solution (\ref{sol-TAG}) satisfies Eq.~(\ref{TAG-eq-1}) automatically, and the electric field is naturally gauge independent. Equivalently, the electric field in (\ref{E-field-TAG}) can be derived from the Maxwell equations (\ref{Maxwell-eqs-x-lin-1}) and (\ref{Maxwell-eqs-x-lin-2}) without any reference to the four-potential $A^\mu$. In contrast, if Feynman-Lorentz gauge is used, current conservation must be explicitly enforced. The authors of \cite{Adil:2006ei} resolved this problem by modifying somewhat artificially the parton's current.

\section{Important configurations}
\label{sec-x0and1}

In this appendix we look at the prolate system in the special case that the wave vector of the unstable mode is perpendicular to the direction of anisotropy (${\bf k}\perp{\bf n}$), and the oblate system when these two vectors are parallel (${\bf k}\parallel{\bf n}$). 
These regions of ${\bf k}$ are important because they are the part of the domain of ${\bf k}$ for which the unstable modes exist up to infinite $k$, see Eqs.~(\ref{k-crit-prolate}) and (\ref{k-crit-oblate}) and Figs. \ref{fig-omega-minus-ex-prolate} and \ref{fig-unstable-modes-oblate}. We will further show that the energy loss is maximal when the velocity of the test parton ${\bf v}$ is parallel to ${\bf n}$ in the prolate plasma and when ${\bf v} \perp{\bf n}$ in the oblate one. The arguments discussed in this appendix are illustrated in Fig. \ref{fig-configurations}. 

We start with the prolate system. The linearized Yang-Mills or Maxwell equations of electric field ${\bf E}(\omega, {\bf k})$ can be
written as  \cite{Carrington:2014bla}
\be
\label{maxwell-1}
\Sigma^{ij}(\omega,{\bf k})E^j(\omega,{\bf k})=0 ,
\ee
with the matrix $\Sigma$ defined by Eq.~(\ref{matrix-sigma}). Since the equation (\ref{maxwell-1}) is homogeneous, there are solutions if the determinant of the matrix $\Sigma$ vanishes - this is the general dispersion equation (\ref{dis-eq-general}). When ${\bf n}= (0,0,1)$ and ${\bf k}=(k,0,0)$, Eq.~(\ref{maxwell-1}) is
\ba
\label{max-pro}
\left[
\begin{array}{ccc}
\omega^2 -  \beta (\omega) 
 & 0 & 0 
\\[1mm]
0 & \omega^2 - k^2 - \alpha(\omega)  & 0
\\[1mm]
0 & 0 & 
\omega^2  - k^2 - \alpha(\omega) - \gamma(\omega)
\end{array}
\right] 
\left[
\begin{array}{c}
E_x(\omega, {\bf k})
\\[1mm]
E_y(\omega, {\bf k})
\\[1mm]
E_z(\omega, {\bf k})
\end{array}
\right] =0 ,
\ea
where
\be
\alpha  (\omega) = \beta (\omega) = \frac{m^2}{2}, 
~~~~~~~~~~
\gamma (\omega) = \frac{m^2 (k^2 -\omega^2)}{2\omega^2} .
\ee
The imaginary modes appear as solutions of the equation $\omega^2  - k^2 - \alpha(\omega) - \gamma(\omega) =0$ which controls the $z$-component of the electric field. Therefore, the exponentially growing component of ${\bf E}$ is parallel to ${\bf n}$. Since the maximal energy loss occurs when ${\bf v} \parallel {\bf E}$, the maximal effect requires ${\bf v} \parallel {\bf n}$.

Let us now consider the oblate plasma. When ${\bf n}= (0,0,1)$ and  ${\bf k}=(0,0,k)$, Eq.~(\ref{maxwell-1}) is 
\ba
\label{max-oble}
\left[
\begin{array}{ccc}
\omega^2 - k^2 - \alpha(\omega)
 & 0 & 0 
\\[1mm]
0 & \omega^2 - k^2 - \alpha(\omega)  & 0
\\[1mm]
0 & 0 & 
\omega^2 - \beta(\omega)
\end{array}
\right] 
\left[
\begin{array}{c}
E_x(\omega, {\bf k})
\\[1mm]
E_y(\omega, {\bf k})
\\[1mm]
E_z(\omega, {\bf k})
\end{array}
\right] =0 ,
\ea
where the coefficients $\alpha(\omega)$ and $\beta(\omega)$ are given by Eq.~(\ref{oblatex1b}). The imaginary modes appear as solutions of the equation $\big(\omega^2  - k^2 - \alpha(\omega) \big)^2 =0$ which controls the $x$-  and $y$-components of the electric field. Therefore, the exponentially growing component of ${\bf E}$ is perpendicular to ${\bf n}$ and the maximal energy loss occurs when ${\bf v} \perp {\bf n}$.


\end{document}